\documentclass[preprint, lefttitle, number, sort&compress]{elsarticle}
\makeatletter
\def\ps@pprintTitle{%
 \let\@oddhead\@empty
 \let\@evenhead\@empty
 \def\@oddfoot{}%
 \let\@evenfoot\@oddfoot}
\makeatother
\usepackage[left=2cm,right=2cm,top=2cm,bottom=3cm]{geometry}
\usepackage{graphicx,import} 
\usepackage{placeins}
\usepackage{subcaption}
\usepackage{natbib}
\usepackage{color}
\usepackage{psfrag}
\usepackage{comment}
\usepackage{array}
\usepackage{multirow}
\usepackage{setspace}
\usepackage[percent]{overpic}
\usepackage{amsfonts}
\usepackage{amsmath}
\usepackage{amssymb}
\usepackage{booktabs}
\usepackage{cases}
\usepackage{caption}
\usepackage{verbatim}
\usepackage[small,compact]{titlesec}
\usepackage{url}
\usepackage[version=4]{mhchem}
\usepackage{textcomp}
\usepackage{flafter}
\usepackage{cleveref}
\usepackage{gensymb}
\usepackage[pagewise]{lineno}
\usepackage[toc, title]{appendix}

\newcommand{\co}{CO$_2$ }

\newcommand{\ts}[1]{\textsuperscript{#1}} 
\newcommand{\of}{OpenFOAM}
\newcommand\rev[1]{{\color{black}#1}} 

\begin{document} 
\begin{frontmatter}
\title{Robustness of point measurements of carbon dioxide concentration for the inference of ventilation rates in a wintertime classroom}
\author[1,2]{Carolanne V.\ M.\ Vouriot\corref{cor1}}
\ead{c.v.vouriot@sheffield.ac.uk}
\author[3]{Maarten van Reeuwijk}
\author[3]{Henry C.\ Burridge}
\cortext[cor1]{Corresponding author}
\address[1]{Department of Civil and Structural Engineering, University of Sheffield, Mappin Street, Sheffield S1 3JD, UK.}
\address[2]{Department of Applied Mathematics and Theoretical Physics, Centre for Mathematical Sciences, University of Cambridge, Wilberforce Rd, Cambridge CB3 0WA, UK.}
\address[3]{Department of Civil and Environmental Engineering, Skempton Building, South Kensington Campus, Imperial College London, London SW7 2BX, UK.}

\begin{abstract}
Indoor air quality in schools and classrooms is paramount for the health and well-being of pupils and staff. Carbon dioxide sensors offer a cost-effective way to assess and manage ventilation provision. However, often only a single point measurement is available which might not be representative of the \co distribution within the room. A relatively generic UK classroom in wintertime is simulated using Computational Fluid Dynamics. The natural ventilation provision is driven by buoyancy through high- and low-level openings in both an opposite-ended or single-ended configuration, in which only the horizontal location of the high-level vent is modified. \co is modelled as a passive scalar and is shown not to be `well-mixed' within the space. Perhaps surprisingly, the single-ended configuration leads to a `more efficient' ventilation, with lower average \co concentration. Measurements taken near the walls, often the location of \co sensors, are compared with those made throughout the classroom and found to be more representative of the ventilation rate if made above the breathing zone. These findings are robust with respect to ventilation flow rates and to the flow patterns observed, which were tested by varying the effective vent areas and the ratio of the vent areas. 
\end{abstract}


\end{frontmatter}
\section{Introduction} \label{sec:intro}
Eleven million pupils and school staff in the United Kingdom spend a significant proportion of their time in school buildings, the majority of which is in classrooms. As such, the indoor air quality of classrooms has the potential to directly affect the health of a large proportion of the population and, indirectly, their families too. Classrooms around Europe have previously been found to experience relatively low levels of ventilation with links being drawn to effects on pupils' health, well-being and academic performance \cite{Haverinen-Shaughnessy2011,Chatzidiakou2012,Salthammer2016,Fisk2017}. The COVID-19 pandemic has brought the issue of indoor air quality in schools and classrooms to the forefront of the attention of the wider public, with airborne disease transmission now a major concern. 

Characterising ventilation rates is critical to understand indoor air quality and building performance, however it can be challenging to do routinely and reliably \cite{Persily2016}. Measuring carbon dioxide (\co\unskip) offers a means to assess a building's ventilation rate for long times, across years, using affordable sensors that are becoming widespread. In classrooms, in the absence of other combustion sources, occupants are the main source of \co through their exhaled breath. High \co concentration is likely to indicate poor ventilation and therefore a likely accumulation of other pollutants, including infected breath. In a single room, \co can also be used as a naturally-occurring tracer gas to measure the ventilation rate, using for instance build up, steady state or decay methods \cite{ISO2017}, all of which have been used in classrooms as reviewed by \citet{Batterman2017} and by \citet{Kabirikopaei2020}. Estimating ventilation rates from \co measurements relies on a number of assumptions \cite{Persily1997,ASHRAE2022}, including: assuming that the \co generation rate is known, which requires knowing the number of occupants and their activity level; as well as, \rev{assuming that each room is a well-mixed single zone so that the \co concentration is uniform and that it can be measured by a single sensor. Previous measurement campaigns in schools and standards have provided differing recommendations for sensor placement \cite{Zhang2022}, an inconsistency that is likely exacerbated by the considerable \co variations measured in classrooms by both \citet{Mahyuddin2014} and \citet{Zhang2022}.} This is especially problematic for naturally ventilated spaces, where the ventilation is not mechanically driven and thermal stratification can be expected to arise. Since the majority of UK classrooms are naturally ventilated, with more than 90\% of energy display certificates issued for school buildings in England and Wales from October 2008 to June 2021 describing their ventilation system as natural ventilation \citep{MHCL}, it is necessary to understand whether a single \co measurement can accurately represent the ventilating flow supplied to the classroom, and by extension, the resulting exposure to contaminants indoors and potential infection risk of the occupants.

In this work, the effect of the ventilating flow on the \co distribution and the resulting accuracy of the estimate of the ventilation rate are assessed using Computational Fluid Dynamics (CFD). In such simulations, the ventilating flow rate can be directly measured and compared to the estimate obtained from \co measurements, which is rarely, if ever, achieved in operational classrooms. In addition, with CFD, the specific effects of the ventilation provision on the \co spatial distribution can be determined without other confounding factors (like sensor accuracy or changes in occupancy levels) which are inevitable in field measurements. Choice is made herein to focus on a steady state method where the \co concentration is assumed to have reached a steady level before calculating the ventilation rate. This choice was made for several reasons, although it is acknowledged that, quite often, in operational classrooms the timescales are such that a steady \co concentration might only be expected to be reached after several hours and so practical observations are challenging. One reason for the choice is that primary school classrooms are the focus and, in the UK, they can be assumed to host school children for longer periods, increasing the chance of reaching a steady state. Considering a steady state scenario does not need a representative timetable for the pupils, which might differ between schools and classrooms. Using a decay method for instance, requires knowledge of when all occupants leave the room and, in addition, assumes that the ventilation rate assessed during unoccupied hours is the same as that during occupied hours, which for naturally ventilated classrooms (typical in the UK) is not likely to be the case \cite{Batterman2017}. Although it can overestimate the ventilation rate \cite{Seppanen1999}, the steady state method is often used in schools and \citet{Kabirikopaei2020} have shown that it is the method with the lowest uncertainty. Assuming the measured peak concentration value to indicate the steady value, as illustrated by \citet{Haverinen-Shaughnessy2011} for instance, also gives an upper bound for the ventilation provision and is useful to identify spaces with insufficient ventilation. Finally, by considering a steady state scenario, the spatial variations in \co concentration can be isolated from transient effects to comprehensibly determine how they are affected by the ventilation provision. For these reasons, we herein simulate rooms in steady state.

The reference scenario considered herein is one driven by horizontal convection which has been observed both experimentally and numerically in spaces with a distributed buoyancy source \cite{Vouriot2023}. This paper aims to assess whether for the reference scenario considered here, in which the ventilation is buoyancy driven through low and high-level vents and convection is assumed to be the dominant source of heat transfer, the \co concentration can be assumed to be well-mixed, and if not then: what is the degree of uncertainty that is introduced by assuming the air in the classroom is well-mixed? To inform the potential variability within a classroom, two limiting ventilation configurations are investigated. Between the two configurations the horizontal location of the top opening on the ceiling is changed: in one configuration, a flow from one end of the classroom to the other is promoted, herein the `opposite-ended' configuration; and in the other, both vents are positioned at one end of the classroom, herein the `single-ended' configuration. The impact of the ventilating flow rate on the results is also assessed, varying the size of the openings to either restrict or enhance the flow, as well as the effect of the ratio of the areas of the high- and low-level vents. In practice, many \co sensors are deployed by attaching them to walls (a viable choice from many perspectives) but the extent to which these measurements might be representative of the \co concentration within the bulk of the space, and particularly in the breathing zone, is not known. As such, this paper also investigates the degree of any additional uncertainties introduced when point measurements of \co are taken near the walls; in addition the sensitivity of the \co measurements to height is also investigated. 

The methodology is introduced in \S\ref{sec:method}. \S\ref{sec:Config} compares the opposite-ended and single-ended configurations for a reference scenario: the resulting ventilating flow is described, the \co concentration in the room and in the breathing zone are compared and the impact of using wall measurements is also discussed In \S\ref{sec:FlowRate}, the reference set-up is modified to investigate the sensitivity of the previous results to changes in the ventilating flow rate by varying the low and high-level vent areas. \S\ref{sec:ventratecalc} details how the inferred ventilation rate is affected by the location of the \co measurement when compared to the other uncertainties associated with estimating the ventilation provision. Finally, conclusions are drawn in \S\ref{sec:conc}. 

\rev{\section{Methodology} \label{sec:method}
The basis of this study are CFD simulations of the Reynolds averaged Navier-Stokes equations within an idealised naturally ventilated classroom during wintertime that were described in detail in \citet{Vouriot2023}. The numerical simulations investigate an idealised model of a classroom which replicates certain aspects broadly representative of a typical naturally ventilated UK classroom and the objective here is to extend these simulations with \co  emissions in order to provide estimates of \co exposure within the expected order of magnitude. A single classroom is simulated with ventilation openings at low- and high-level (\Cref{fig:ClassroomDimensions}), and the heat losses at the walls, air leakage, radiative effects and thermal mass are neglected. The effects of the openings are considered only through the ventilating flow they provide. These openings are modelled as flat open surfaces on the floor and ceiling of the room, therefore also avoiding the consideration of bi-directional flows through the ventilation openings. Using floor and ceiling openings is expected to be broadly representative of most buoyancy dominated ventilation flows with high- and low-level openings, irrespective of their precise orientation \cite{Linden1990}. We note that beyond buoyancy-dominated ventilation flows, wind can have significant effect, either opposing the flow \citep{Hunt1999} or enhancing mixing \citep{CarrilhodaGraca2018}, but the incorporation of these effects remain the focus of future research. 

A wintertime scenario is considered as it corresponds to the time of year during which classrooms might be least ventilated and have been shown to exhibit the highest concentrations of \co \citep{Vouriot2021}. Although classrooms are subject to a wide variety of heat sources, we pursue a lumped approach here and apply all heat inputs (heaters, people, solar gains, electronic devices) to the floor area. Following similar reasoning, the \co input is also approximated by the addition of a passive tracer uniformly distributed over the classroom's cross-section, which is deemed appropriate to densely occupied spaces such as classrooms, and released steadily at breathing height. This, again, is a simplification since the release of human breath is both buoyant and periodic. These effects are important in the near-field, but the focus of this study is the far-field where the injection method is less important. 

Initially, two ventilation configurations are investigated. In each of these, the opening areas, heat input and therefore overall flow rate are kept constant but the position of the top vent is changed leading to:
\begin{itemize}
\item an opposite-ended configuration with vents at opposite ends of the room, and 
\item a single-ended configuration with vents on the same end of the room. 
\end{itemize}
The heat input from radiators in the room is adjusted to enable the provision of a comfortable thermal environment to the pupils and staff. The focus is on a primary school classroom as younger children are more vulnerable to indoor contaminants and therefore a more at risk population. The occupants of the classroom are assumed to be of primary school age (between the ages of 5 and 11 in the UK) when defining their \co and heat outputs. The occupants' breathing is modelled as a steady volumetric source of \co\unskip, represented by the addition of a continuous planar release of a passive scalar at breathing height (set here to be between 1.1 and 1.2\,m, the typical breathing height of seated primary school pupils from the BB101 guidance \cite{BB101}). 

Following these assumptions, the exact reference scenario to be simulated is selected following the available guidance. The classroom is assumed to have 32 occupants following the BB101 guidelines \citep{BB101}. The classroom is sized using the minimum requirements given by BB103 \citep{BB103} and also corresponds to the reference case of \citet{Jones2021}: the surface area is set to 55\,m\ts{2}, with a ceiling height of 2.7\,m. An illustration of the set-up considered is shown in \Cref{fig:ClassroomDimensions}. Ventilation is driven through high- and low-level openings located on the floor and ceiling of the classroom. In the initial scenario, the low-level vent has an area $A_l$ of 0.4\,m\ts{2} (0.8\,m $\times$ 0.5\,m) centred at coordinates (1.4, 1.25, 0). Two ventilation configurations are then investigated by changing the location of the upper-level vent, while keeping its area $A_h$ constant and equal to 0.2\,m\ts{2} (0.5\,m $\times$\,0.4\,m). In the opposite-ended (OE) configuration, it is positioned at the opposite end of the classroom from the low-level opening and centred around (8.75, 4.3, 2.7). In the single-ended (SE) configuration, the opening is positioned on the same end as the bottom opening and centred at coordinates (1.25, 4.3, 2.7). The 32 occupants in the room are each assumed to produce 60\,W of heat \citep{BB101} and a \co generation rate of $G = 3.35\times 10^{-6}$\,m\ts{3}/s, which is the average for primary aged school children aged between 6 and 11 as reported by by \citet{Persily2017}, assuming an activity level of 1.4\,met (which corresponds to an increase in the child's activity proportional to the increase in activity level of an office worker writing or standing, compared to the levels for rest). We do not explicitly account for the slightly higher \co generation of the one, or two, teaching staff likely to be present in the room --- the impact of such differences falls within the uncertainty of the \co generation rate; for example, due to differences in activity level within the occupants. In order to be precise for any particular group of occupant, their generation rate would have to be representative of their level of activity, age and gender, all of which can lead to large variations between individuals. Since the focus on this paper is not on the effect of the different occupants' position on the \co distribution, a consistent individual generation rate $G$ is used and distributed uniformly across the room. An additional 4,280\,W of heat is supplied at the floor to take into account the classroom heating provision and ensure a comfortable room temperature. The ambient outside temperature $T_a$ is taken to be 5\degree C, assumed to be typical for the coldest months in the UK. 

\begin{figure}[h!]
    \centering
 	 \def\svgwidth{\textwidth}
 	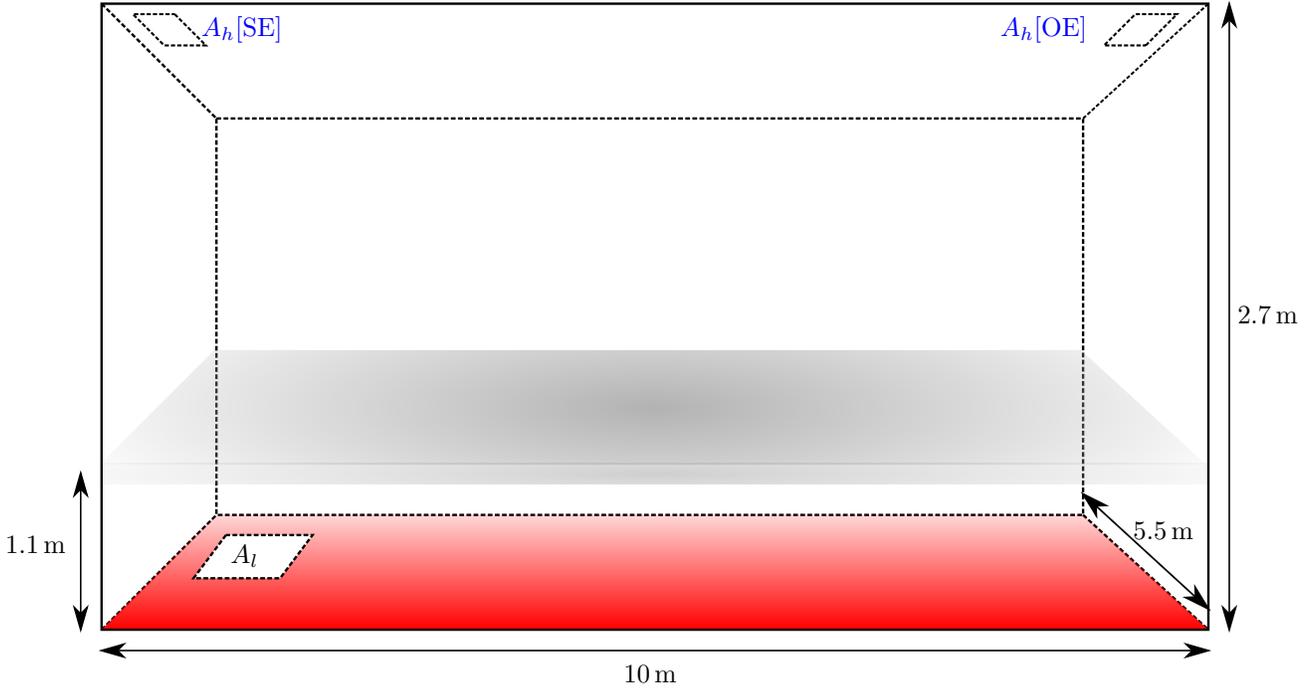
  \caption{Illustration of the set-up used to represent a generic naturally ventilated UK classroom (either in a single-ended (SE) or opposite-ended (OE) configuration). The heat is input on the floor as shown in red here. A passive scalar representing \co is introduced at breathing height (between 1.1 and 1.2\,m).}
    \label{fig:ClassroomDimensions}
\end{figure}

\begin{table}[h] 
    \centering
     \begin{tabular}{|c|c|c|c|}
     \hline 
    Parameter & Symbol & Unit & Input \\
    \hline
    Floor surface area & $S_c$ & m\ts{2} & 55.0  \\ 
    Classroom volume& $V_c$ & m\ts{3} & 148.5  \\ 
    Classroom height & $H_c$ & m & 2.7 \\
    Number of occupants & $N$ & - & 32 \\ 
    Inlet area & $A_l$ & m\ts{2} & 0.40  \\ 
    Outlet area & $A_h$ & m\ts{2} & 0.20  \\
    Heat input & $W_c$ & W & 6200  \\ 
    \co generation rate & $G$ & m\ts{3}/s & $3.35\times10^{-6}$\\
    Outdoor temperature & $T_a$ & \degree C& 5 \\
    Outdoor \co concentration & $C_a$ & ppm & 400 \\
    Ambient density & $\rho_a$ & kg/m\ts{3} & 1.268 \\
    Thermal expansion coefficient & $\alpha$ & $K^{-1}$ & 0.00362 \\
    Heat capacity & $c_p$ & J/kg/K & 1005 \\
    \hline 
     \end{tabular}
     \caption{Inputs used to define the classroom simulation.}
     \label{tab:SetupInput}
 \end{table}
 
The CFD simulations are run using \of{}v2106, developed by OpenCFD \citep{OFv2106}. A transient solver, \texttt{buoyant\-Pimple\-Foam}, is used due the presence of long-time period fluctuations, potentially characteristic of internal wave modes, within the data generated by steady solvers despite the absence of time-dependent or varying boundary conditions. To reduce run-times, our simulations are initialised with a stratification established by a steady run. Subsequently, the simulations are run with a transient solver for 8,600\,s with statistics averaged only over the last 3,600\,s. These values are chosen to ensure a suitable convergence and accuracy of the results while limiting computational costs. Overall, each simulation took approximately 15 hours to run (using 1,024 cores) on the UK National Supercomputing Service's ARCHER2 (i.e. each simulation consumed around 15,360 cpu hours). 

Simulations are run using the $k-\omega$ SST turbulence model. This model strikes a balance between the performance of $k-\epsilon$ models for shear flows and $k-\omega$ models at the walls \citep{Pope2000}, both features relevant to the capturing ventilating flows. Large variations in the turbulence levels in the classroom can also be expected \citep{Blocken2018}, leading to potential relaminarisation of the flow which the $k-\omega$ SST model is expected to capture. The chosen \of{} version also incorporates the effects of buoyancy on turbulence production by using the \texttt{buoyancy\-Turb\-Source} finite volume option \citep{OFbuoyancy}. 

The computational domain includes the classroom along with two exterior boxes linked to the classroom through inlets and outlets, included in order to properly model the effect of flow at the two openings and the resulting flow in the classroom, as shown in \Cref{fig:Mesh}. The classroom is defined as a cube of dimensions 10\,m $\times$ 5.5\,m $\times$ 2.7\,m. The bottom exterior box is centred around the inlet to the classroom, with dimensions 2.8\,m $\times$ 2.5\,m $\times$ 1\,m. The top exterior covers the same surface as the classroom with dimensions 10\,m $\times$ 5.5\,m $\times$ 3.7\,m. The classroom is linked to the exterior boxes via an inlet and an outlet of the same size as the vents with respective areas $A_l$ and $A_h$, and height 0.2\,m. The sensitivity of the simulations to the size of the external boxes was tested and these dimensions were chosen as they did not impact the results. The mesh is defined as a perfect orthogonal mesh with $\Delta x=\Delta y =\Delta z =0.05 $\,m in the classroom, inlet and outlet and $\Delta x=\Delta y =\Delta z =0.1 $\,m in the exterior, leading to 1,399,356 hexahedral cells. A grid convergence study was also performed: the mesh used in this study accurately represents both the bulk flow and \co distribution when compared to a finer mesh (with $\Delta x=\Delta y =\Delta z =0.025 $\,m), full details are presented by \citet{Vouriot2022} and \citet{Vouriot2023}. 

\begin{figure}
    \centering
    \def\svgwidth{\textwidth}
\begingroup%
  \makeatletter%
  \providecommand\color[2][]{%
    \errmessage{(Inkscape) Color is used for the text in Inkscape, but the package 'color.sty' is not loaded}%
    \renewcommand\color[2][]{}%
  }%
  \providecommand\transparent[1]{%
    \errmessage{(Inkscape) Transparency is used (non-zero) for the text in Inkscape, but the package 'transparent.sty' is not loaded}%
    \renewcommand\transparent[1]{}%
  }%
  \providecommand\rotatebox[2]{#2}%
  \newcommand*\fsize{\dimexpr\f@size pt\relax}%
  \newcommand*\lineheight[1]{\fontsize{\fsize}{#1\fsize}\selectfont}%
  \ifx\svgwidth\undefined%
    \setlength{\unitlength}{2170.5bp}%
    \ifx\svgscale\undefined%
      \relax%
    \else%
      \setlength{\unitlength}{\unitlength * \real{\svgscale}}%
    \fi%
  \else%
    \setlength{\unitlength}{\svgwidth}%
  \fi%
  \global\let\svgwidth\undefined%
  \global\let\svgscale\undefined%
  \makeatother%
  \begin{picture}(1,0.55390034)%
    \lineheight{1}%
    \setlength\tabcolsep{0pt}%
    \put(0,0){\includegraphics[width=\unitlength,page=1]{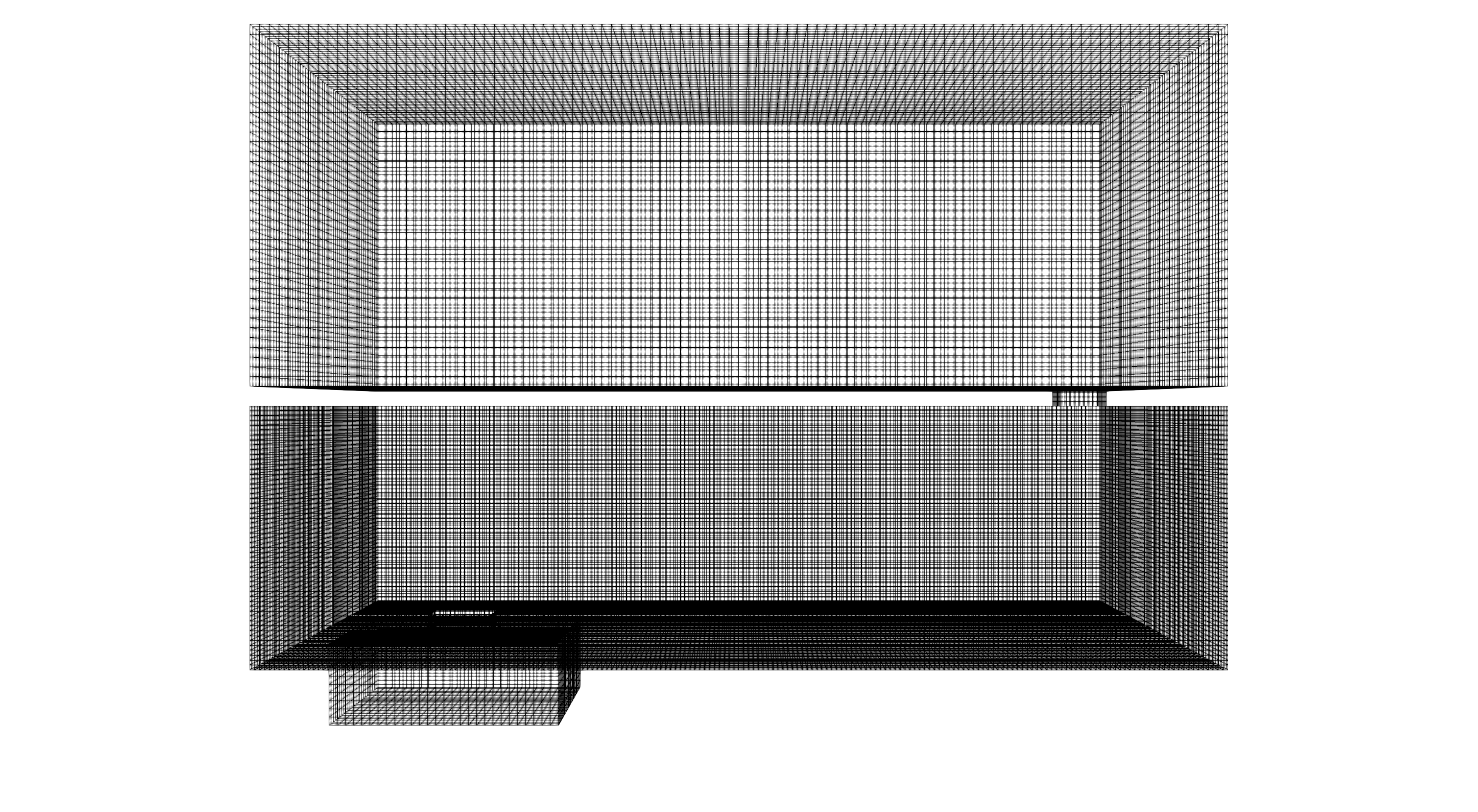}}%
    \put(0.05365074,0.18380028){\color[rgb]{0,0,1}\makebox(0,0)[lt]{\lineheight{1.25}\smash{\begin{tabular}[t]{l}Classroom\end{tabular}}}}%
    \put(0.03194944,0.40321901){\color[rgb]{0,0,1}\makebox(0,0)[lt]{\lineheight{1.25}\smash{\begin{tabular}[t]{l}Top exterior\end{tabular}}}}%
    \put(0.09740784,0.13130189){\color[rgb]{0,0,1}\makebox(0,0)[lt]{\lineheight{1.25}\smash{\begin{tabular}[t]{l}Inlet\end{tabular}}}}%
    \put(0.40473958,0.0696423){\color[rgb]{0,0,1}\makebox(0,0)[lt]{\lineheight{1.25}\smash{\begin{tabular}[t]{l}Bottom exterior\end{tabular}}}}%
    \put(0.84560962,0.27463286){\color[rgb]{0,0,1}\makebox(0,0)[lt]{\lineheight{1.25}\smash{\begin{tabular}[t]{l}Outlet\end{tabular}}}}%
    \put(0,0){\includegraphics[width=\unitlength,page=2]{Mesh.pdf}}%
    \put(0.41935368,0.54415497){\color[rgb]{0,0,1}\makebox(0,0)[lt]{\lineheight{1.25}\smash{\begin{tabular}[t]{l}Domain outflow\end{tabular}}}}%
    \put(0.23060796,0.04266518){\color[rgb]{0,0,1}\makebox(0,0)[lt]{\lineheight{1.25}\smash{\begin{tabular}[t]{l}Domain inflow\end{tabular}}}}%
  \end{picture}%
\endgroup%

    \caption{Computational grid used to simulate the classroom, including the modelled exterior, inlet and outlet.}
    \label{fig:Mesh}
\end{figure}

At all walls, including the classroom's and the exterior boxes' (apart from the domain inflow and outflow), a no-slip velocity boundary condition is used. All walls are also assumed to be adiabatic, with the exception of the floor where a constant 6,200\,W heat input is imposed. The scalar representing the breath of occupants through the addition of \co is added over the surface area of the classroom between the heights of 1.1 and 1.2\,m with a source corresponding to the generation rate $NG$. At the domain inflow the temperature and scalar are set to ambient and at the domain outflow a Neumann boundary condition is used setting the gradient of the temperature and scalar fields to 0. Velocity boundary conditions at the inflow and outflow are calculated from the pressure field where the pressure difference $\Delta p_0 = -\rho_a g H_{domain}$ is imposed across the domain inflow and outflow. In addition, the bulk flow rate and room temperature were checked against the well-mixed predictions of \citet{Gladstone2001} \rev{(described in Appendix \ref{ap:wellmixed})} and the flow pattern arising from the distributed heat source was comparable to what was observed in small-scale experiments \cite{Vouriot2023}.}
\FloatBarrier
\section{Comparing the opposite-ended and single-ended configurations}\label{sec:Config}

The results of our simulations show the ventilation flow patterns differ between the opposite-ended and single-ended configurations. We report on these differences, before then evidencing the robustness of our results to different ventilation rates and the distribution of vent areas in \S\ref{sec:FlowRate}.

\subsection{Bulk quantities}\label{sec:flowComparison}
The simulated ventilating flow rate, mean temperature, and \co concentration in the classroom are given for both configurations in \Cref{tab:Config_results}. The effective area $A^*$ is 0.18 and if the discharge coefficient is taken to be identical at each vent, $c_d = c_l = c_h$, it is close to 0.7, which lies within the experimentally determined range of $0.6 \leq c_d \leq 1.0$ \citep{Heiselberg2001,Gladstone2001}. The well-mixed predictions are presented in square brackets taking the value for the discharge coefficients $c_l$ and $c_h$ determined in Appendix \ref{sec:CdCalc}. 

\begin{table}[h!]
\centering
     \begin{tabular}{|c|c|c|c|c|cc|cc|c|}
          \cline{3-10} 
          \multicolumn{2}{c|}{} & $Q$ & ACH & $Q_p$ & \multicolumn{1}{c|}{$\Delta T$} & $\Delta T_{e}$ & \multicolumn{1}{c|}{$\Delta C$} & $\Delta C_{e}$ & $\eta$\\ 
          \multicolumn{2}{c|}{} & (m\ts{3}/s) & (-)& (L/s/person) & \multicolumn{1}{c|}{(\degree C)}&  (\degree C) & \multicolumn{1}{c|}{ (ppm)} & (ppm)&  $(-)$ \\ 
          \hline 
          \multicolumn{2}{|c|}{\multirow{2}{*}{Opposite-ended}}& 0.239  & 5.85 & 7.54 &  19.3 &  20.4 & 427&  449 & 1.05  \\ 
          \multicolumn{2}{|c|}{} & [0.243] & [5.90] & [7.60] & \multicolumn{2}{c|}{[20.0]} & \multicolumn{2}{c|}{[441]} & [1]\\ 
          \hline 
          
          \multicolumn{2}{|c|}{\multirow{2}{*}{Single-ended}}& 0.241  & 5.85 & 7.54 &  18.9 &  20.2 & 357&  444 & 1.24  \\ 
          \multicolumn{2}{|c|}{} & [0.247] & [5.98] & [7.71] & \multicolumn{2}{c|}{[19.7]} & \multicolumn{2}{c|}{[435]} & [1]\\ 
          \hline 
     \end{tabular}
    \caption{Results of the CFD simulations for the opposite-ended and single-ended configurations, and well-mixed predictions (shown in square brackets) using the loss coefficient at the vents $c_h$ and $c_l$ calculated in Appendix \ref{sec:CdCalc}, for: the ventilating flow rate, the room averaged excess temperature $\Delta T = {T_c}-T_a$ and excess \co concentration $\Delta C = {C_c}-C_a$. In all five cases, $T_e$ and $C_e$ correspond to the exhaust values calculated from the measured ventilating flow rate $Q$ assuming a well-mixed environment, the corresponding excess temperature and \co concentration are calculated from $\Delta T_e = T_e -T_a$ and $\Delta C_e = C_e -C_a$ respectively. $\eta$ gives a measure of the contaminant removal efficiency.}
    \label{tab:Config_results}
\end{table}

The bulk flow rates attained from the CFD simulations for both the opposite-ended and single-ended configurations are very similar, differing by only 1\%. These bulk flow rates agree with the predictions from the well-mixed model to within about 3\%, an appropriate accuracy for almost all applications. For both cases, $Q_p$, the flow rate per occupant, is close to 8\,L/s/person, in line with the BB101 guidelines \citep{BB101}, which advises ventilation rates of 8--9\,L/s/person to achieve \co concentrations of 1,000\,ppm in a typical classroom. 

\Cref{tab:Config_results} shows that there is only a slight difference in the room averaged excess temperature $\Delta T$ between the two configurations. Unsurprisingly, both values are quite close to the well-mixed predictions as expected given the similarities in flow rate and enforcing conservation of energy in the steady state. In addition, temperature is an active tracer such that variations in a horizontal plane result in buoyancy forces which act to promote mixing. Conversely, for the case of the room-averaged excess \co concentration $\Delta C$ (which is a passive tracer within the simulations), the difference between the two simulations is around 16\%. Interestingly, the average \co concentration in the single-ended configuration is \textit{lower} than in the opposite-ended configuration. This indicates, perhaps counter-intuitively, that if one takes \co concentration as an indoor air quality indicator, then the single-ended ventilation could be considered as having better indoor air quality. We investigate this finding in more detail below. 

The differences between the two configurations, and the extent to which \rev{the well-mixed predictions presented in Appendix \ref{ap:wellmixed} are appropriate}, are investigated by calculating the expected exhaust temperature $T_{e}$ by rearranging \cref{eq:Qpred} and \cref{eq:Tpred}, and \co concentration $C_{e}$ from \cref{eq:CO2pred}, using the flow rate $Q$ obtained numerically. This enables the contaminant removal efficiency, $\eta$, to be examined --- this metric has frequently been used to analyse the performance of ventilation configurations \citep[e.g.][]{Novoselac2003,CarrilhodaGraca2016}, and we define
\begin{equation}
    \eta = \frac{\Delta C_e}{\Delta C} = \frac{C_{e}-C_a}{{C_c}-C_a} \;.
    \label{eq:efficiencyEta}
\end{equation}
Although similar to the ventilation efficiency or effectiveness defined by \citet{Sandberg1981} and used, for instance, by \citet{ASHRAE2002}, here the contaminant is not introduced at the inlet but generated within the room. If $\eta$ is equal to unity, the efficiency is equivalent to the one found in a perfectly mixed room. Values below unity indicate short-circuiting and values superior to unity are representative of displacement ventilation systems \citep{CarrilhodaGraca2016}. For the single-ended simulation, the contaminant removal efficiency is found to be equal to 1.24 and indicative of the performance of displacement ventilation. Practically, this implies that if the room averaged \co concentration $C_c$ were used to predict the ventilation rate, the ventilation rate would be overestimated, in this case by 24\%. In the opposite-ended configuration, although still superior to unity (with $\eta = 1.05$), $\eta$ is a lot closer to the efficiency expected in a perfectly well-mixed room and indicative of a less effective ventilation configuration than the single-ended configuration. 

\subsection{Flow pattern and vertical variations}
\begin{figure}[h!]
    \begin{center}
    \begin{subfigure}[b]{\textwidth}
            \centering
         \includegraphics[width=\textwidth]{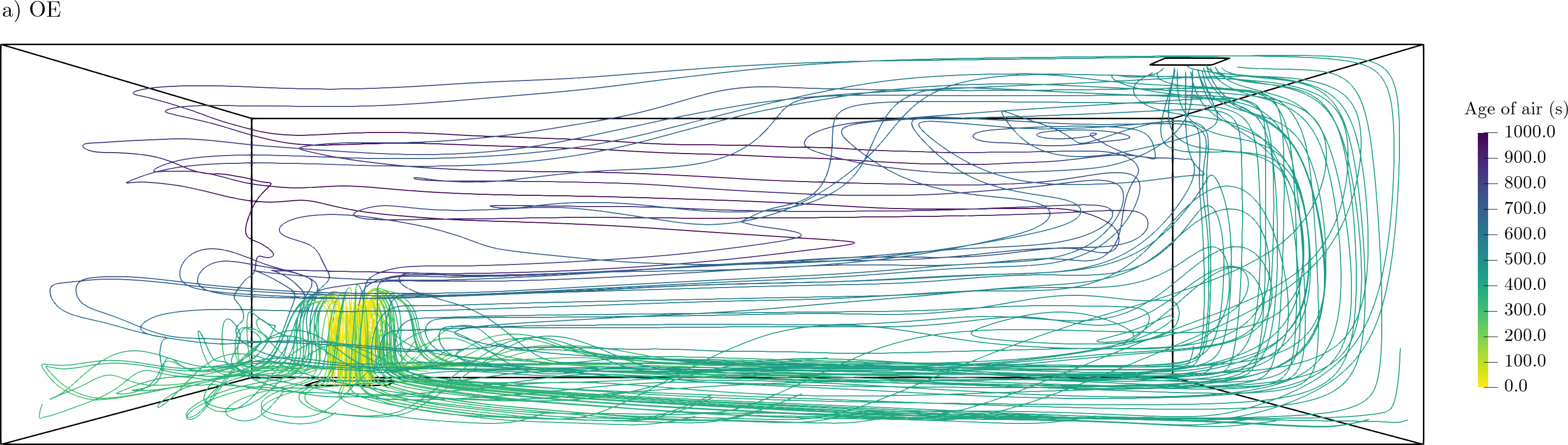}
    \vspace{0.05cm}
     \end{subfigure}
     
     \begin{subfigure}[b]{\textwidth}
         \centering
         \includegraphics[width=\textwidth]{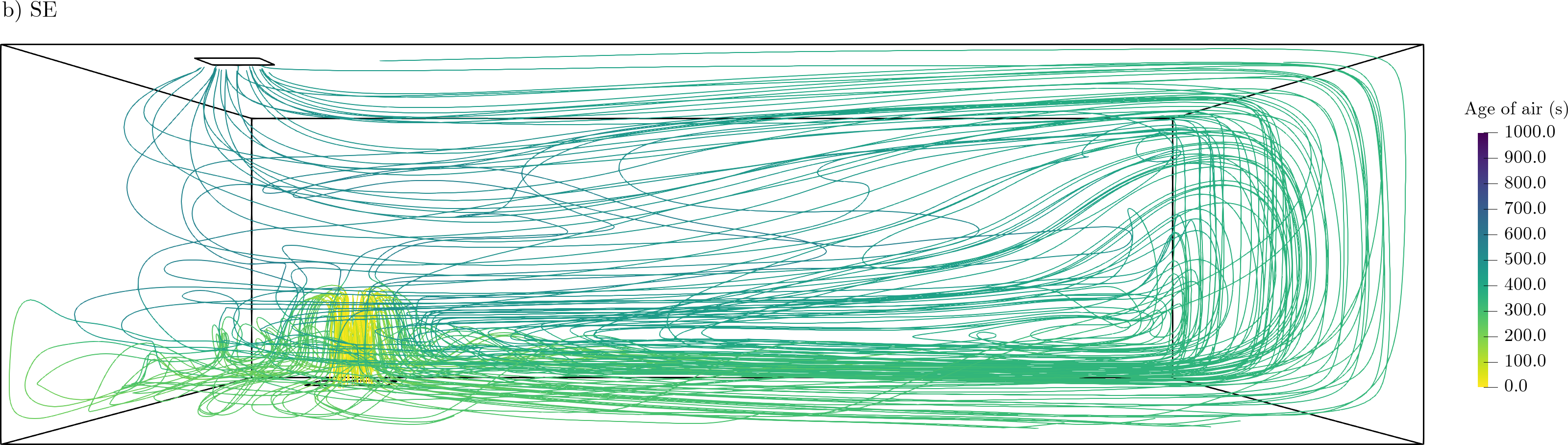}
    \end{subfigure}
    \end{center}
    \caption{Streamlines originating from the low-level vent in both a) the opposite-ended and b) single-ended configurations coloured by the age of air. In both cases, 100 streamlines are seeded at the classroom inlet.}
    \label{fig:ConfigStreamlines}
    \end{figure}

To better understand the factors underlying the difference in performance of the two configurations, the flow pattern is investigated via the study of the streamlines within the classroom. \Cref{fig:ConfigStreamlines}\unskip a) and \Cref{fig:ConfigStreamlines}\unskip b) show the pattern of the ventilating flow for the opposite-ended and single-ended configurations, respectively. In each case, the streamlines are coloured by the age of air --- a statistic which represents, the time that air at a given point in the room has been within the room, and hence gives a measure of the `freshness' of the air. 

In both configurations, the incoming cold air can be seen flowing into the classroom through the low-level vent, forming a fountain \citep{Hunt2015}, and falling back down to the floor. This relatively cold air then flows along the floor across the room before hitting the opposite wall and flowing upwards. In the opposite-ended configuration (\Cref{fig:ConfigStreamlines}\unskip a), the flow partially leaves the classroom directly through the high-level vent. In the single-ended configuration however (\Cref{fig:ConfigStreamlines}\unskip b), the air then flows along the ceiling before reaching the classroom outlet on the other side of the classroom. \Cref{fig:ConfigStreamlines}\unskip b) shows that the single-ended configuration does not lead to short-circuiting, and, on the contrary, creates a ventilating flow that covers a large proportion of the classroom. On the other hand, \Cref{fig:ConfigStreamlines}\unskip a) shows a large area of the room away from the outlet in the opposite-ended configuration in which there is little ventilating flow. This area contains relatively stale air, and likely leads to an accumulation of \co concentration, therefore increasing the overall room average. This is of concern for exposures since this region of stagnating air is located within the breathing zone. This investigation of the streamlines, and associated age of air, sheds light on the differences in the flow patterns between the two configurations which give rise to, and were evidenced by, the differences in the bulk metrics for each configuration as shown by \Cref{tab:Config_results}.

\begin{figure}[h!]
    \centering
    \includegraphics[width =\textwidth]{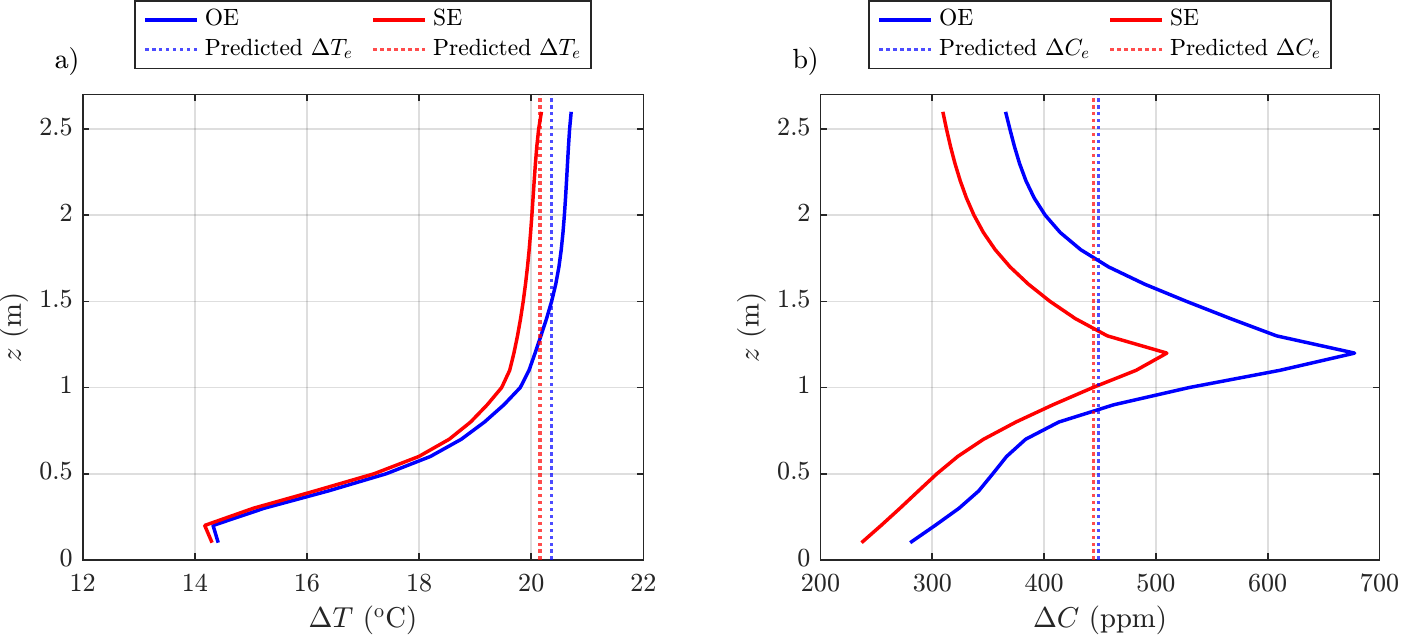}
    \caption{Horizontally averaged excess: a) temperature and b) \co concentration in the classroom for the opposite-ended (OE) and single-ended (SE) configurations. $\Delta T_{e}$ and $\Delta C_{e}$ are the predicted excess exhaust temperature and \co concentration calculated from the measured flow rate. }
    \label{fig:ConfigProfileComp}
\end{figure}

The differences in the ventilating pattern between the two configurations is also visible in the horizontally averaged excess temperature and \co concentration (see \Cref{fig:ConfigProfileComp}). The horizontally averaged excess temperature (\Cref{fig:ConfigProfileComp}\unskip a) increases with height, from about 14\degree C near the floor to over 20\degree C near the ceiling. Below a height of 0.75\,m, both ventilation configurations exhibit notionally identical temperatures, but higher up the opposite-ended configuration exhibits elevated horizontally averaged excess temperatures, by around 1\degree C. Defining the heat removal efficiency as $\Delta T_e/\Delta T$ (equivalent to the definition of $\eta$), results in values of 1.06 for the opposite-ended and 1.07 for the single-ended configurations, which are close to a well-mixed environment. The differences in removal efficiency of \co between the configurations, compared to the one determined for temperature, are likely due to one being a passive and the other an active scalar, as discussed previously. 

The similarity in temperature profiles is in contrast to \Cref{fig:ConfigProfileComp}\unskip b) which shows the horizontally averaged excess \co concentration in the classroom. Both configurations exhibit a significant variation in \co concentration with height which broadly follow a similar profile shape in which the \co concentration peaks within the breathing zone (1.0\,m--1.5\,m) where \co is introduced. However, the \co concentration is higher in the opposite-ended scenario, than the single-ended, at all heights. We note that whilst the opposite-ended configuration contaminant removal efficiency is found to be close to a well-mixed value ($\eta=1$), this occurs in the presence of significant vertical variations in the horizontally averaged \co concentration around the well-mixed value, it just so happens that in this case the variations approximately sum to zero --- there is no requirement that this is inherently the case, as illustrated by the findings for the single-ended configuration. 

\begin{figure}[h!]
    \centering
    \includegraphics[width =\textwidth]{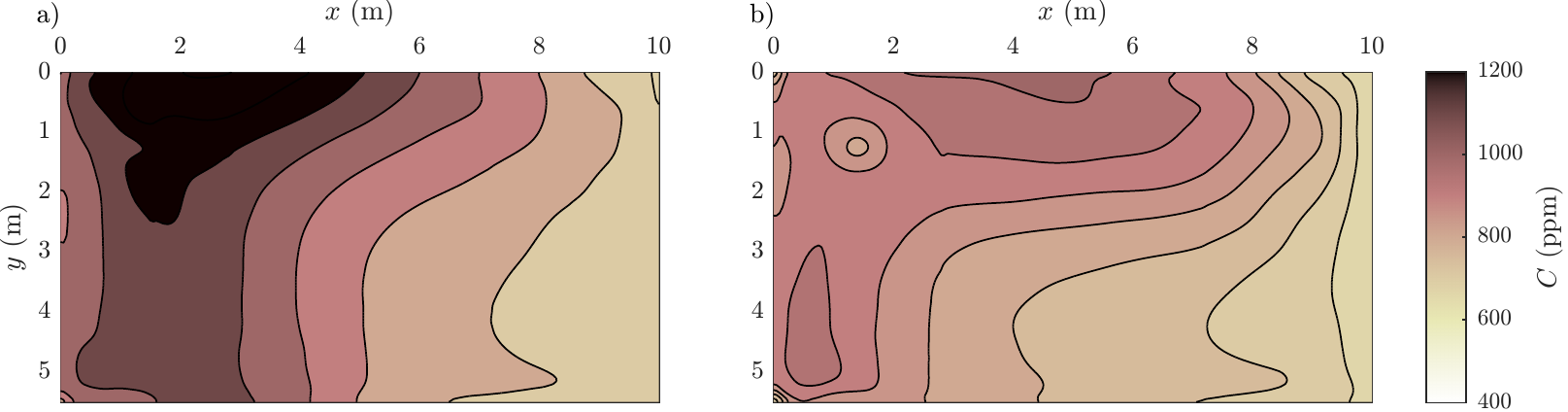}
    \caption{Horizontal cross section of the \co concentration, vertically averaged, in the breathing zone ($1\leq z \leq 1.5$\,m) for: a) the opposite-ended and b) the single-ended configuration. }
    \label{fig:ConfigCO2bz}
\end{figure}

\Cref{fig:ConfigCO2bz} shows the \co concentration variation across the room, averaged over the height of the breathing zone. Both configurations, exhibit significant variations around the well-mixed \co concentration ($\approx 840$\,ppm). In particular, for the single-ended configuration, the concentration is shown to vary from around 700 to 1,400\,ppm, with the lowest concentration found in the region of the rising flow (see \Cref{fig:ConfigStreamlines}). This is also observed in \Cref{fig:ConfigCO2bz}\unskip b), for the single-ended configuration, although the \co variations are lower, only spanning the range 700 to 1,000\,ppm. This difference in the distribution of \co in the breathing zone can be attributed to the different ventilating patterns (\Cref{fig:ConfigStreamlines}) and the mechanisms underlying these different patterns are discussed by \citet{Vouriot2023}. 
\FloatBarrier

\subsection{\co spatial distribution}
Given the significant differences between the two ventilation configurations, and its relevance as an indicator of indoor air quality, the \co distribution is examined more closely in each ventilation configuration by looking at a wider variety of statistics. \Cref{tab:Config_BZresults} compares the \co distribution in the breathing zone ($1\leq z \leq 1.5 $\,m) and the overall classroom, by reporting values for the mean \co concentration and the coefficient of variation in each region. The coefficient of variation of the \co concentration, $cVar$, is calculated in each domain by dividing the unbiased estimate of the standard deviation by the mean. 
\begin{table}[h!]
    \centering
     \begin{tabular}{|c|c|c|c|c|}
          \hline 
          Configuration & Location & $C$ (ppm) & $cVar$ (\%) \\ 
          \hline 
          Opposite-ended & Classroom &  827 &  16.6  \\ 
          & Breathing zone &  986 &  18.7  \\ 
          \hline 
          Single-ended & Classroom &  757 &  11.5  \\ 
          & Breathing zone &  852 &  11.7  \\ 
          \hline 
     \end{tabular}
    \caption{Statistics of the \co concentration for the overall classroom and the breathing zone ($1\leq z \leq 1.5 $\,m) in the two ventilation configurations. ${C}$ is the mean \co concentration in each domain and the coefficient of variation $cVar$ is the ratio of the unbiased estimate of the standard deviation relative to the mean.}
    \label{tab:Config_BZresults}
\end{table}
The mean \co concentration in the breathing zone is higher than for the overall classroom by 13\% for the single-ended configuration, and 20\% for the opposite ended. This increase is not unexpected given that the \co is generated in this zone. What may be notable when looking at occupants' exposures, is that in the breathing zone for the opposite-ended configuration, the mean \co is 17\% (about one standard deviation) higher than in the single-ended configuration, and the coefficient of variation is also significantly increased.

\begin{figure}[t!]
    \centering
    \includegraphics[width =\textwidth]{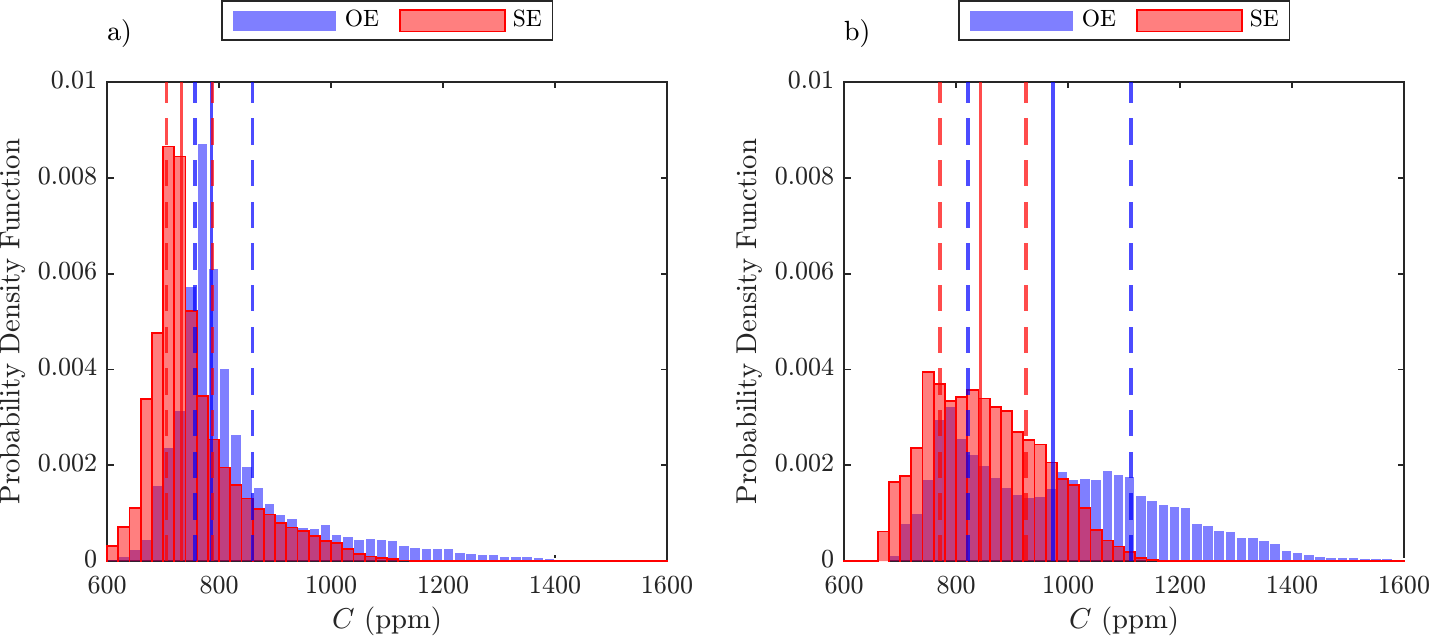}
    \caption{Histogram of the \co concentration for the opposite-ended and single-ended configuration for: a) the whole classroom and b) the breathing zone ($1\leq z\leq 1.5$\,m) using 50 bins. For each case, the median is shown with a full vertical line, the first and third quartiles are shown with dashed vertical lines.}
    \label{fig:Configpdf}
\end{figure}

Histograms of the distribution of \co in both configurations are also shown in \Cref{fig:Configpdf}: a) for the entire classroom, and b) in the breathing zone only. For reference the median values and interquartile ranges are marked by solid and dashed vertical lines respectively. All four distributions are skewed with large tails towards higher \co values and in both regions, higher values of \co are present in the opposite-ended configuration. \Cref{fig:Configpdf}\unskip b) evidences that the values within the breathing zone are significantly elevated and are bimodally distributed in the case of the opposite-ended configuration. The breadth of the \co distributions, particularly in the breathing zone, highlight potential challenges in representing or estimating exposures with a limited number of point measurements, typical of existing sensing technologies. 

\subsection{Sensor placement}\label{sec:Configsensor}

\begin{figure}[h!]
    \centering
    \includegraphics[width =\textwidth]{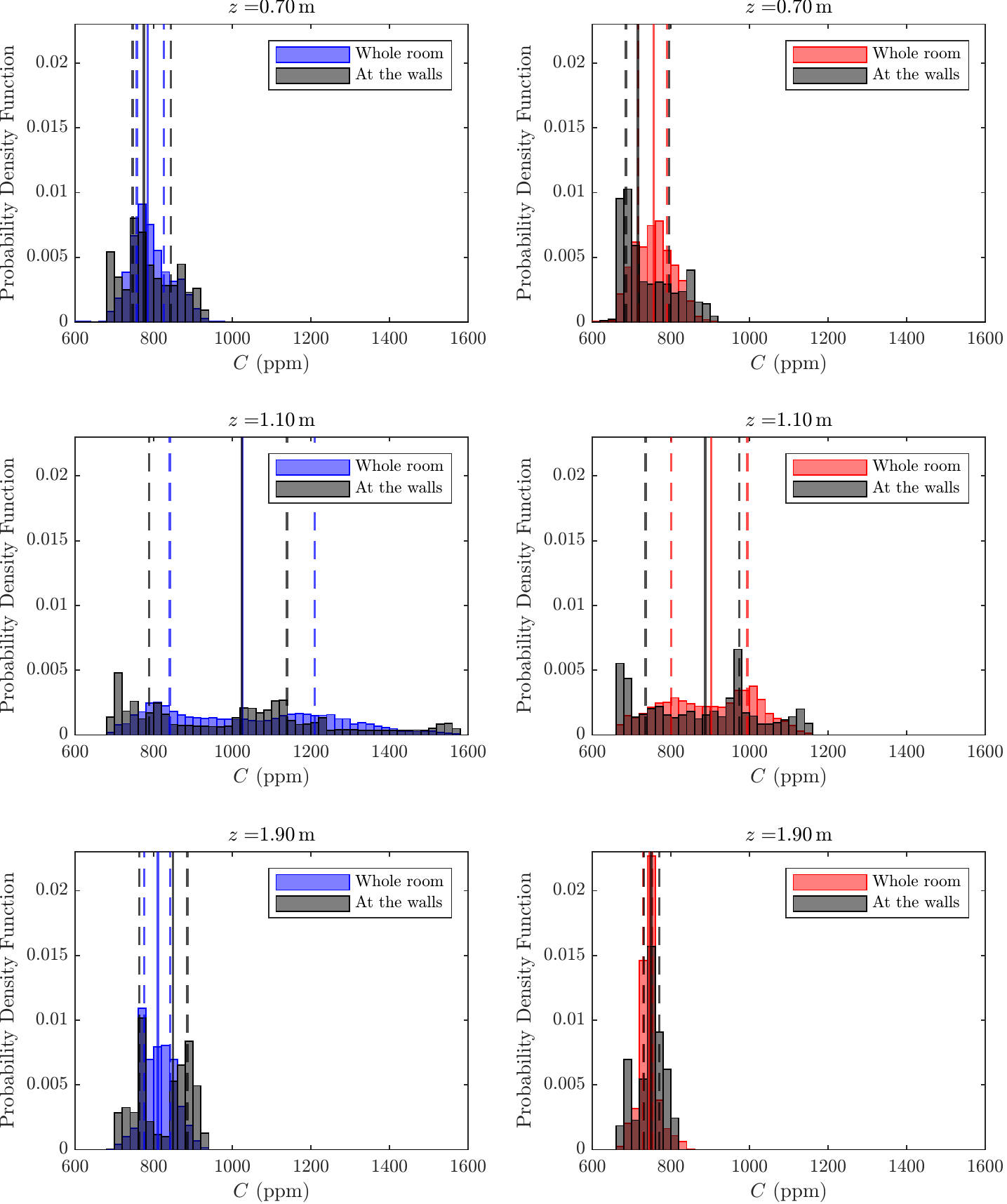}
    \caption{Histogram of the \co concentration for the opposite-ended (left) and single-ended configurations (right) at three heights using 50 bins. The \co concentration across the whole room at every height is compared to measurements at the walls (defined as within 0.2\,m of a wall). For each case, the median is shown with a full vertical line, the first and third quartiles are shown with dashed vertical lines.}
    \label{fig:ConfigpdfWall}
\end{figure}

The simulations presented here provide access to the full three dimensional fields in the classroom; we exploit this to assess where \co sensors can be placed in order to take measurements of greater relevance to occupants. In particular, we analyse the \co concentrations within the room as a whole and compare these to the concentrations measured only near the walls, since sensors placement is often restricted to walls --- in our case taken to be within a region 0.2\,m from the walls. Histograms of the \co distribution at different heights are plotted for the opposite-ended and single-ended configurations (\Cref{fig:ConfigpdfWall}). The greatest variation is observed in the breathing zone, e.g. $z=1.1$\,m. At the lower heights shown, differences are evident between the \co concentrations in that plane and those measured at the walls. Although the mean and median values typically agree quite well, differences in some of the interquartile values highlight the changes in the distributions. However, since the flow enters at low-level and exits at high-level, and, as the flow evolves it can only mix, in both of the configurations examined, measurements above the breathing zone result in greater agreement between measurements in the plane and at the walls. If \co measurements at the walls are to be used to estimate a ventilating flow rate, measurements taken above the breathing zone are found to be more likely to give accurate readings. This is increasingly true at heights above 1.9\,m, and it can be assumed that this result is likely to hold for ventilation designs promoting inflows at low-level and outflows at high-level as long as measurements are never made above the height of the high-level outlet vent(s). 

\begin{table}[h!]
    \centering
     \begin{tabular}{|c|c|c|c|c|}
          \hline 
          Simulation & At height & $\epsilon$ (\%)& $cVar$ (\%)&  $cVar_{w}$ (\%)\\ 
          \hline 
          Opposite-ended & $z=1.1$\,m & 2.9 &  20.4 &  23.9  \\ 
           & $z=1.9$\,m & 1.7 &  5.1 &  8.1  \\ 
          \hline 
          Single-ended & $z=1.1$\,m & 2.5 &  13.0 &  16.7  \\ 
          & $z=1.9$\,m & 0.4 &  3.3 &  4.7  \\ 
          \hline 
     \end{tabular}
    \caption{Accuracy of the \co measurements at the walls at two heights for the two ventilation configurations. $\epsilon$ is the error in the predicted mean when considering the \co concentration near the walls (defined as within 0.2\,m of a wall). $cVar$ is the coefficient of variation of the \co concentration over the whole room at a given height and $cVar_{w}$ quantifies the coefficient of variation at the same height considering measurements near the walls only.}
    \label{tab:Config_Wallresults}
\end{table}

\Cref{tab:Config_Wallresults} summarises these results by comparing statistical results at heights 1.1\,m and 1.9\,m. The wall-error, $\epsilon$, is first calculated to quantify the error in the average predictions made based on measurements in the wall region compared with the average of the cross section at that height, namely
\begin{equation}
    \epsilon = \frac{|\langle C_w\rangle-\langle C\rangle|}{\langle C\rangle}\,,
\end{equation} 
where $\langle C_w\rangle$ is the mean \co concentration at a given height measured near the walls and $\langle C\rangle$ is the mean \co concentration across the whole room at that height. The table shows that whilst the difference in the wall-error is approximately doubled when measurements are made at $z=1.1$\,m, the wall-error is small, i.e. less than the measurement error for typical low cost non-dispersive infrared sensors. However, the coefficients of variations highlight that since a low number of (often single) point measurements are made within a given space, then, for these types of ventilation designs, measurements made higher up are more prudent.
\FloatBarrier
\section{Sensitivity of the ventilating flow and indoor environment to different
ventilation rates and distribution of vent areas}\label{sec:FlowRate}
In this section, a range of different vent areas and vent area ratios are selected to investigate how robust our results might be. The different simulations selected to perform this analysis and are summarised in \Cref{tab:ResultsFlowRate}. The reference set-up analysed above (\S \ref{sec:Config}) is included and described as the original vent set-up (denoted, `OS'). The effective areas of the classroom openings are varied: they are reduced in simulations SO and SOR, and enlarged in the LO and LOR configurations. In the smaller openings (SO) set-up, both the inlet and outlet areas are halved and they are both doubled in the larger openings (LO) case --- thus keeping the vent area ratio, $A_l/A_h$, unchanged. This is not the case for the SOR and LOR configurations, where only the area of the high-level vent is changed to achieve effective vent areas $A^*$ equivalent to those calculated for the SO and LO cases respectively (\textit{a priori} needing to assume that the discharge coefficients remain unchanged). The SOR case corresponds to a scenario with smaller openings where the vent area ratio is halved and the LOR case is a set-up with larger openings where the vent area ratio is quadrupled.

\subsection{Bulk quantities and flow pattern}
The bulk flow rates, room averaged temperatures and \co concentrations within the classroom are given in \Cref{tab:ResultsFlowRate} for each simulation and ventilation configuration. Broadly, the flow rates obtained are close to that predicted using the well-mixed predictions summarised in square brackets in \Cref{tab:ResultsFlowRate} and calculated using the discharge coefficient $c_l$ and $c_h$ (given for each case in Appendix \ref{sec:CdCalc}). Single-ended configurations consistently lead to larger flow rates, in accordance with the discussion in \S \ref{sec:Config}; these differences are reduced (sometimes unnoticeable to the two decimal places reported) for cases with smaller openings that result in lower ventilation flow rates (SO and SOR).

\begin{table}[h!]
\centering
     \begin{tabular}{|c|c|c|c|c c|c c|c c|c|c c|c|c|c|}
     \hline 
     \multirow{2}{*}{Case} & \multirow{2}{*}{Vents}  & \multirow{2}{*}{$A_h/A_l$}& $A^*$& \multicolumn{2}{c|}{\multirow{2}{*}{ACH}} & \multicolumn{2}{c|}{$Q_p$} & \multicolumn{2}{c|}{$\Delta T$} & $\Delta T_{e}$& \multicolumn{2}{c|}{$\Delta C$} & $\Delta C_{e}$ & \multirow{2}{*}{$\eta$} & $L_M$ \\ 
      &  & & (m$^2$)& \multicolumn{2}{c|}{} & \multicolumn{2}{c|}{(L/s/pers.)} & \multicolumn{2}{c|}{(\degree C)} & (\degree C)& \multicolumn{2}{c|}{(ppm)} & (ppm) & & (m)\\ 
     \hline 
     \multirow{2}{*}{OS} & OE & \multirow{2}{*}{0.50} & 0.18 & 5.79 & [5.90] & 7.47 & [7.60] & 19.3 & [20.0] & 20.4 & 427 & [441] & 449 & 1.05 & 0.56 \\ 
      & SE &  & 0.18 & 5.85 & [5.98] & 7.54 & [7.71] & 18.9 & [19.7] & 20.2 & 357 & [435] & 444 & 1.24 & 0.57 \\ 
     \hline 
     \multirow{2}{*}{SO} & OE & \multirow{2}{*}{0.50} & 0.09 & 3.73 & [3.78] & 4.81 & [4.87] & 30.4 & [31.2] & 31.6 & 687 & [687] & 696 & 1.01 & 0.49 \\ 
      & SE &  & 0.09 & 3.73 & [3.79] & 4.81 & [4.88] & 30.2 & [31.1] & 31.6 & 568 & [686] & 696 & 1.23 & 0.49 \\ 
     \hline 
     \multirow{2}{*}{SOR} & OE & \multirow{2}{*}{0.25} & 0.10 & 3.99 & [4.06] & 5.15 & [5.23] & 28.1 & [29.1] & 29.5 & 761 & [640] & 651 & 0.86 & 0.32 \\ 
      & SE &  & 0.10 & 4.00 & [4.07] & 5.15 & [5.25] & 28.0 & [29.0] & 29.5 & 546 & [639] & 650 & 1.19 & 0.32 \\ 
     \hline 
     \multirow{2}{*}{LO} & OE & \multirow{2}{*}{0.50} & 0.34 & 8.88 & [9.15] & 11.45 & [11.79] & 12.2 & [12.9] & 13.3 & 279 & [284] & 293 & 1.05 & 0.63 \\ 
      & SE &  & 0.35 & 8.94 & [9.26] & 11.52 & [11.93] & 11.9 & [12.7] & 13.2 & 246 & [281] & 291 & 1.18 & 0.64 \\ 
     \hline 
     \multirow{2}{*}{LOR} & OE & \multirow{2}{*}{2.00} & 0.32 & 8.54 & [8.83] & 11.01 & [11.39] & 12.5 & [13.4] & 13.8 & 246 & [294] & 304 & 1.24 & 1.00 \\ 
      & SE &  & 0.33 & 8.66 & [8.92] & 11.16 & [11.50] & 12.5 & [13.2] & 13.6 & 227 & [291] & 300 & 1.32 & 1.02 \\ 
     \hline 
     \end{tabular}  
    \caption{Bulk flow and room averaged results for the different simulations, the simulations described above correspond to the original vent set-up (OS). $A^*$ is calculated from the flow rate and integral of buoyancy obtained in the simulations following \cref{eq:AstarCFD}. Predicted values using the calculated values for $c_l$ and $c_h$ (given in Appendix \ref{sec:CdCalc}) are shown in square brackets. $\Delta T_e$ and $\Delta C_e$ are calculated based on the measured flow rate assuming a well-mixed environment as described in \S\ref{sec:Config}. $\eta$ is a measure of the contaminant removal efficiency. $L_M$ is the jet-length, characteristic of the rise height of the fountain that forms at the low-level vent }
    \label{tab:ResultsFlowRate}
\end{table}

In general, a larger flow rate leads to lower mean \co concentration; this is just one factor that leads to higher values for the opposite-ended configuration in all cases. The contaminant removal efficiency $\eta$ metric is calculated leading to similar results to the original set-up (OS) for the smaller and larger openings cases where the vent area ratio is kept the same (SO and LO); i.e. a value close to unity for the opposite-ended configuration and a value close to 1.2 for the single-ended configuration. For the cases where the vent area ratio is altered (SOR and LOR), however, this is not the case --- this is despite similar flow rates to those of the smaller and larger openings cases (SO and LO), respectively, being achieved with identical effective areas. In the opposite-ended configuration, the set-up with smaller openings and vent area ratio (SOR) has an efficiency well below unity which indicates the increased presence of stagnation zones in the classroom and, further, should the room averaged \co concentration be used to predict the ventilation provision, the rate would be underestimated. For the set-up with larger openings and vent area ratio (LOR), on the other hand, the values of ventilation efficiency are elevated, in both configurations above 1.2. The flow structure that arises in both cases is discussed below but already this analysis highlights that the effective area $A^*$ and the heat load are not always adequate to predict appropriate estimates of the mean \co and resulting contaminant removal efficiency; instead, in our case, at least the vent area ratio must also be accounted for.

\begin{figure}[h!]
    \centering
    \includegraphics[width = \textwidth]{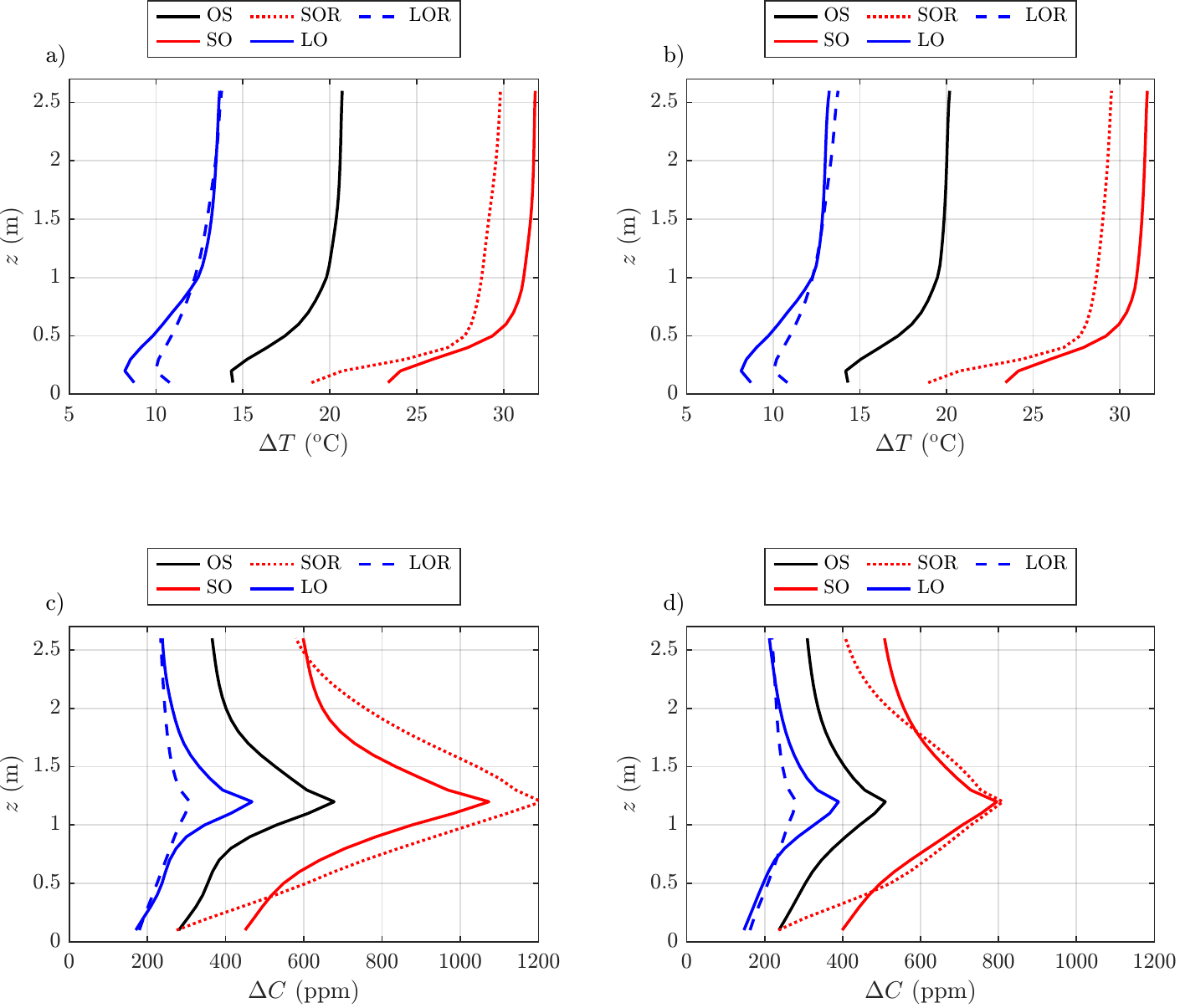}
    \caption{Comparison of the horizontally averaged excess temperature (top panes) and \co concentration (bottom panes) in the classroom for different vent set-ups using the opposite-ended (left panes) and single-ended (right panes) configurations.}
    \label{fig:FRProfileComp}
\end{figure}
 The horizontally averaged excess temperatures and \co concentrations are plotted in \Cref{fig:FRProfileComp}. Trends are similar to those observed in the original vent set-up (OS) for the \co profiles shown in \Cref{fig:FRProfileComp}\unskip c) and \Cref{fig:FRProfileComp}\unskip d), with a significant increase in \co concentration observed for the different opposite-ended configurations when compared to the respective single-ended set-ups. Two notable exceptions are observed. Firstly, the \co concentration in both simulations in the set-up with larger openings and vent area ratio (LOR) varies significantly less across the height of the classroom than what is observed for the simulations with larger openings and the original vent area ratio (LO); although the concentration is similar low down ($z \lesssim 0.8$\,m) and high up ($z \gtrsim 2.2$\,m), the \co concentration in the breathing zone is considerably lower in the simulations where the vent area ratio is increased (LOR) when compared to those where it is kept the same as the original set-up (LO). Secondly, the \co concentration in the simulations with smaller openings and vent area ratio (SOR) peaks above all other simulations within the breathing zone but the concentration near the floor is smaller than that of some other scenarios. Qualitatively the temperature profiles shown in \Cref{fig:FRProfileComp}\unskip a) are close to those in \Cref{fig:FRProfileComp}\unskip b), i.e. those for the opposite-ended configurations are broadly similar to those of the single-ended configurations. As expected, increasing the vent areas leads to a decrease in the room temperature (clearly indicated by the room averaged temperature in \Cref{tab:ResultsFlowRate}) in the set-ups with larger openings (LO and LOR). However, due to the increased vent area ratio, the temperature in the lower third of the room is higher in the LOR case than for the larger openings case with the original vent area ratio (LO). In the smaller openings cases (SO and SOR), on the other hand, changing the vent area ratio has an effect over the entire height of the room, with lower temperatures found in the set-up where the vent area ratio is halved (SOR) when compared to the one with original vent area ratio (SO). 

\begin{figure}[h!]
\vspace{-1cm}
    \begin{center}
        \begin{subfigure}[b]{\textwidth}
            \centering
         \includegraphics[width=\textwidth]{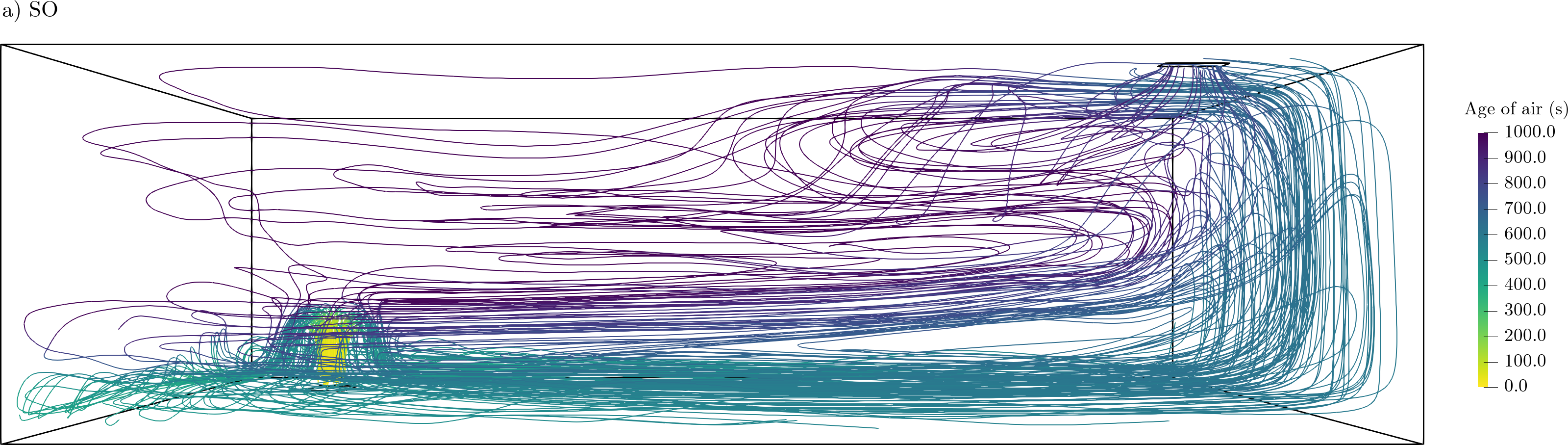}
         \label{fig:StreamlinesSO1OE}
     \end{subfigure}
     ~
    \begin{subfigure}[b]{\textwidth}
            \centering
         \includegraphics[width=\textwidth]{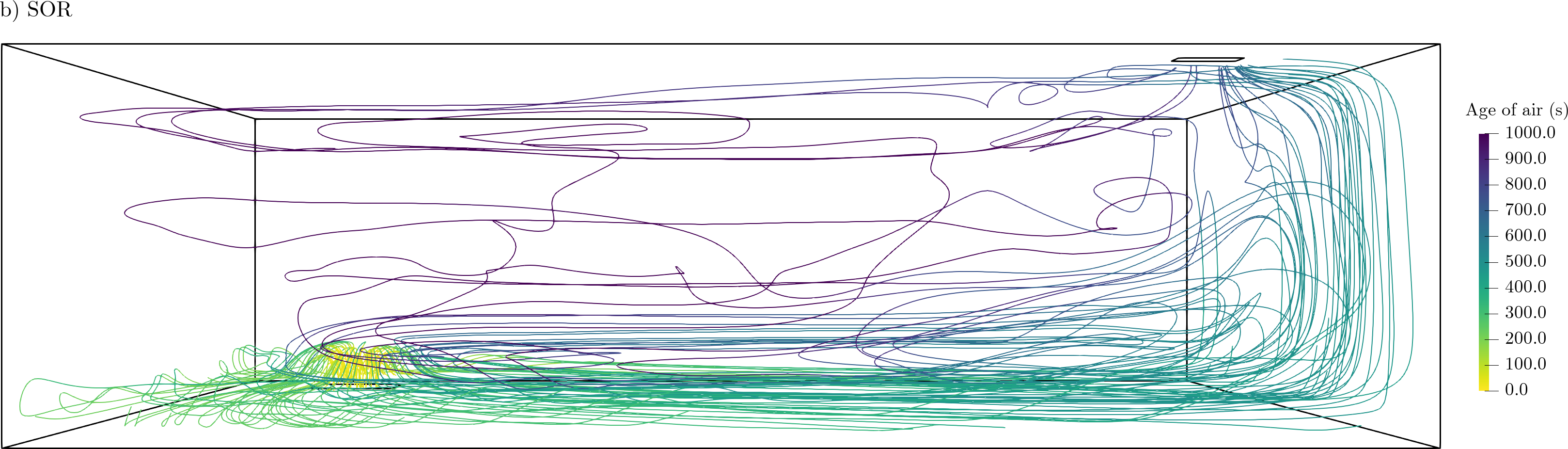}
         \label{fig:StreamlinesCFSO2}
     \end{subfigure}
     ~
        \begin{subfigure}[b]{\textwidth}
            \centering
         \includegraphics[width=\textwidth]{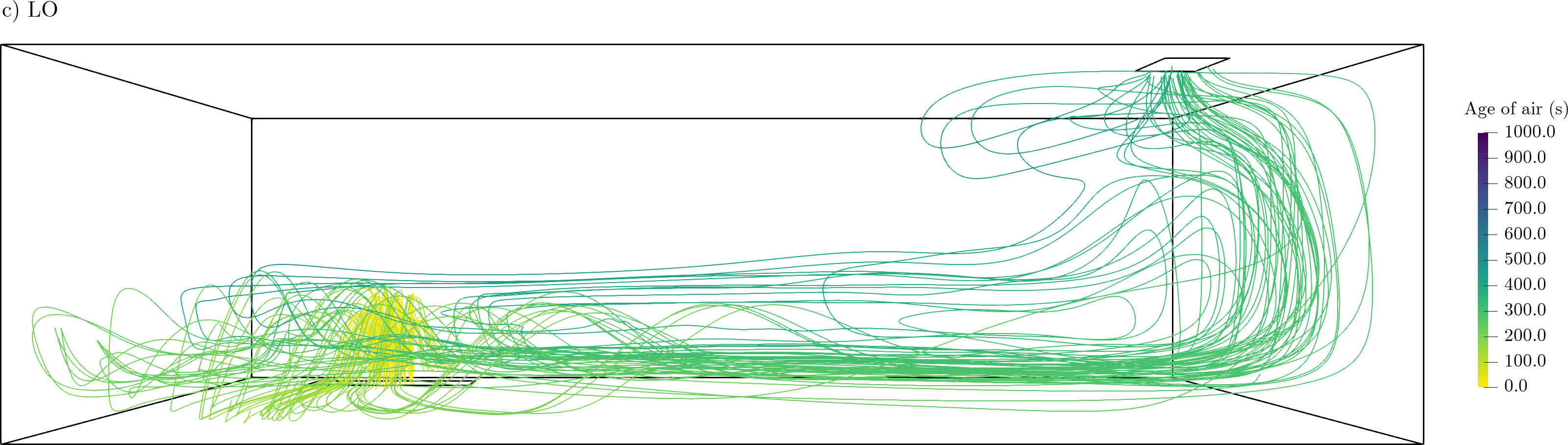}
         \label{fig:StreamlinesLO1OE}
     \end{subfigure}
     ~
     \begin{subfigure}[b]{\textwidth}
         \centering
         \includegraphics[width=\textwidth]{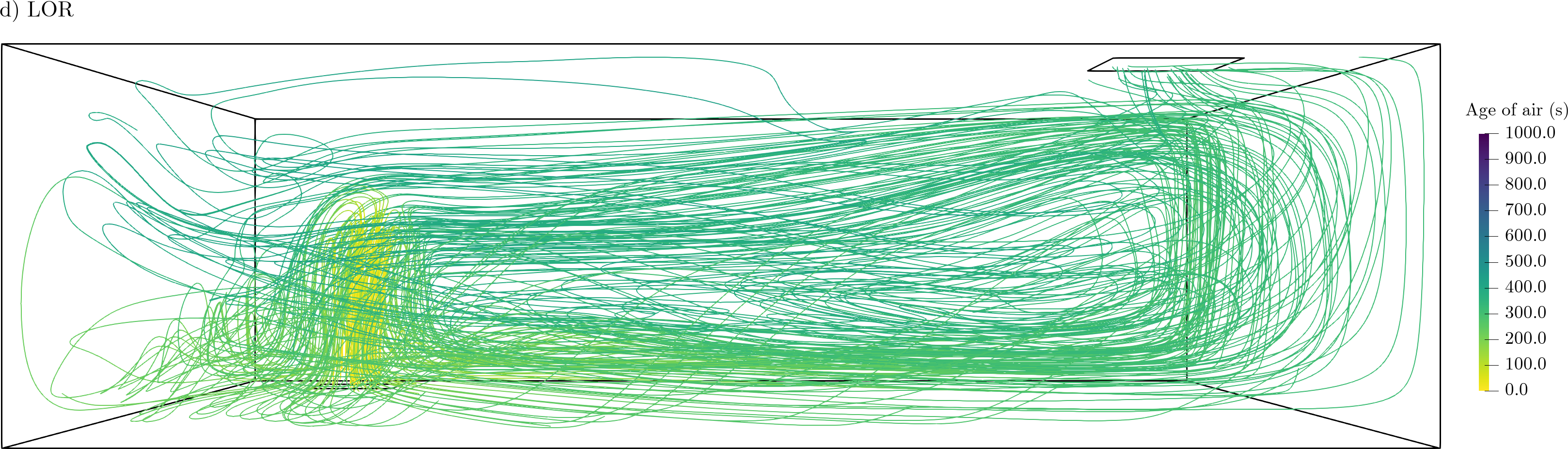}
         \label{fig:StreamlinesCFLO2}
    \end{subfigure}
    \end{center}
    \caption{Streamlines originating from the low-level vent in the opposite-ended configuration for a) the SO, b) SOR, c) LO and d) LOR cases coloured by the age of air. In both cases, 100 streamlines are seeded at the classroom inlet.}
    \label{fig:Streamlines2}
    \end{figure}

The ventilation patterns in the opposite-ended configuration for the set-ups where the vent area ratio is modified (SOR and LOR) are studied further by looking at the streamlines in each case plotted in \Cref{fig:Streamlines2} and compared to the equivalent smaller and larger openings set-ups with the original vent area ratio (SO and LO). These can be directly compared to \Cref{fig:ConfigStreamlines}\unskip a), which shows the streamlines for the original vent set-up (OS), also in the opposite-ended configuration. OS, SOR and LOR have identical inlets and heat input, their only difference being the area of the high-level vent leading to different vent area ratios and effective areas. By keeping the low-level vent area constant, the effect of the change in ventilating flow rate can be seen directly. Some of the differences in the mixing that arise within the classroom can be examined through the lens of the mixing induced by the fountain that forms as cooler air rises from the lower vent. For such flows the mixing induced is known to scale with products of the source volume flux, in our case the ventilation flow rate, and powers of the source Froude number $Fr$ \citep{Burridge2016}, see Appendix \ref{sec:CdCalc} for details including the values attained in each simulation. Moreover, the vertical extent over which such mixing is induced is constrained by the rise height of the fountain which is known to be a product of the physical length scale of the source and powers of the source Froude number \citep{Burridge2012}, or in the simplest case their linear product, i.e. the jet-length. As discussed by \citet{Hunt2015}, these findings hold for fountains formed at rectangular sources, such as the vents in our simulations and so we define jet-length as $L_M = \sqrt{A_l} Fr$. With a low flow rate, in the case with smaller openings and vent area ratio (SOR) case, the fountain rises significantly less than the original vent (OS) or the smaller openings set-up with the original vent area ratio (SO), as expected based on the shorter jet-length, and this creates a much larger stagnating zone, with very stale air as indicated by the old age of air and higher \co concentrations observed in \Cref{fig:FRProfileComp}. In the case with larger openings and vent area ratio (LOR) however, the increase in ventilating flow rate also increases the height of the fountain at the inlet which rises more than halfway across the height of the classroom, much higher than the set-up with similar larger openings but with the original vent area ratio (LO) although the ventilating flow rate is comparable. This encourages recirculation and mixing within the room and no clear stagnation zones are visible. Instead, the streamlines cover most of the classroom and the air is a lot fresher overall, leading to age of air values close to 400\,s, half of what can be observed in the original vent set-up (OS) and the set-up with smaller openings and vent area ratio (SOR), and further explains the low \co concentrations observed in \Cref{fig:FRProfileComp}. For both cases where the vent area ratio is different from the original vent set-up (SOR and LOR) considered here, the contaminant removal efficiency is useful to identify cases which do not match previous patterns and which require further investigation. 
 
\subsection{\co spatial distribution}

\begin{figure}[h!]
    \centering
    \includegraphics[width = \textwidth]{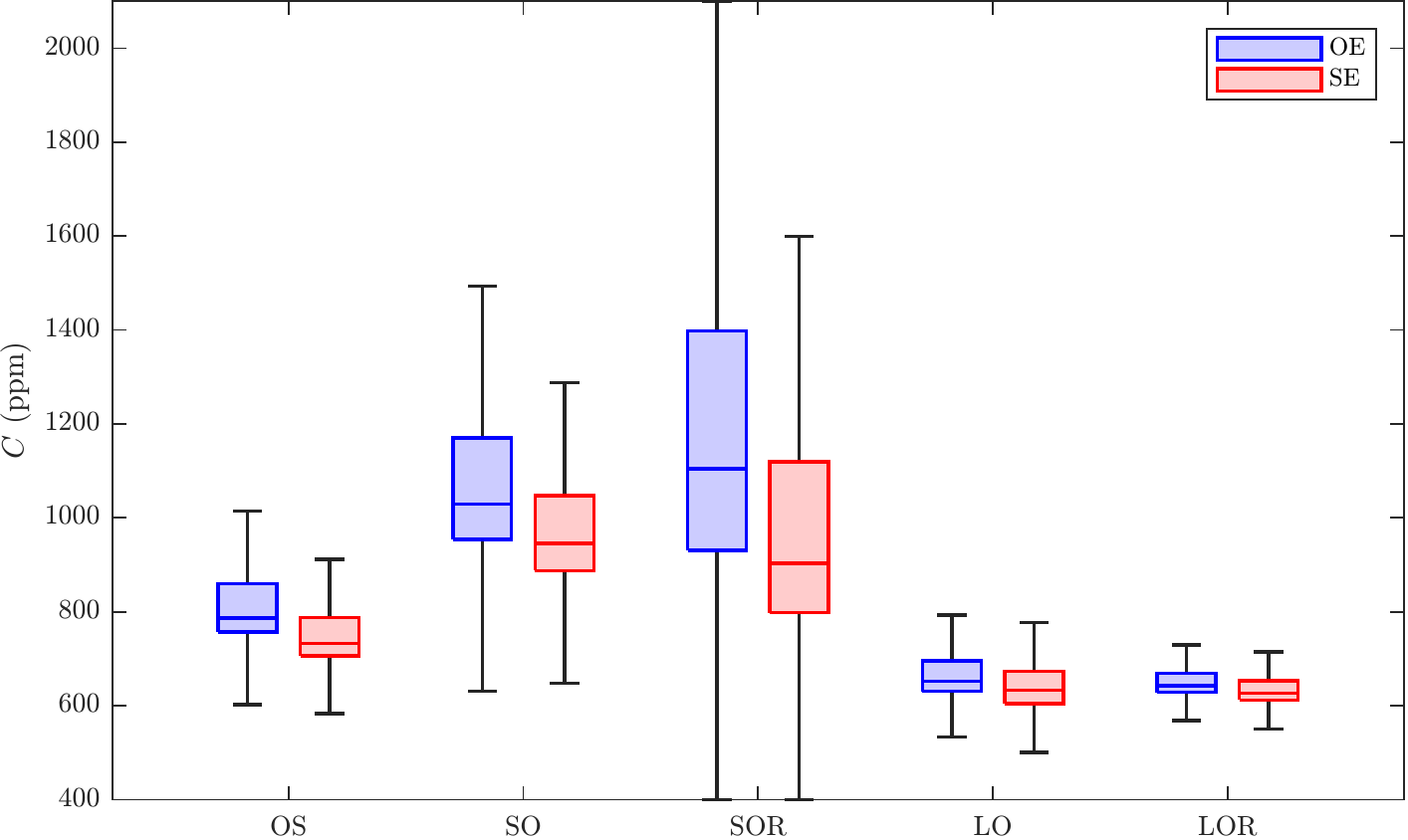}
    \caption{\co distribution in the classroom for different vent set-ups in the opposite-ended (blue) and single-ended (red) configurations. The bottom and top of the box show the 25\ts{th} and 75\ts{th} percentiles, the median is shown by the central horizontal line. The whiskers include data within one and a half of the interquartile range from the 1\ts{st} and 3\ts{rd} quartiles.}
    \label{fig:FRBoxCO2}
\end{figure}

The impact of the changes in scenario on the \co concentration in the room is summarised by the box plots shown in \Cref{fig:FRBoxCO2}. This plot compares the classroom \co distribution in each vent set-up for both the opposite-ended and single-ended configurations. 

\Cref{fig:FRBoxCO2} shows that the values of \co in the classroom are consistently lower in the single-ended configurations, often accompanied by lower spreads in the measured values. For similar geometries, the spread and overall \co in the room increases as the flow rate decreases. The differences between the opposite-ended and single-ended distributions reduce as the ventilating flow rate increases, better agreement is seen for instance for the set-ups with larger openings (LO and LOR). As identified above, the case with smaller openings and vent area ratio (SOR) displays the highest spread in \co concentrations, with values ranging from the ambient 400\,ppm to over 2,000\,ppm in the opposite-ended configuration, although its median is similar to the the other set-up with smaller openings (SO). The increased mixing observed for the case with larger openings and vent area ratio (LOR) is also demonstrated with a smaller spread in \co concentrations.

\subsection{Sensor placement}

\begin{figure}[h!]
    \centering
    \includegraphics[width = \textwidth]{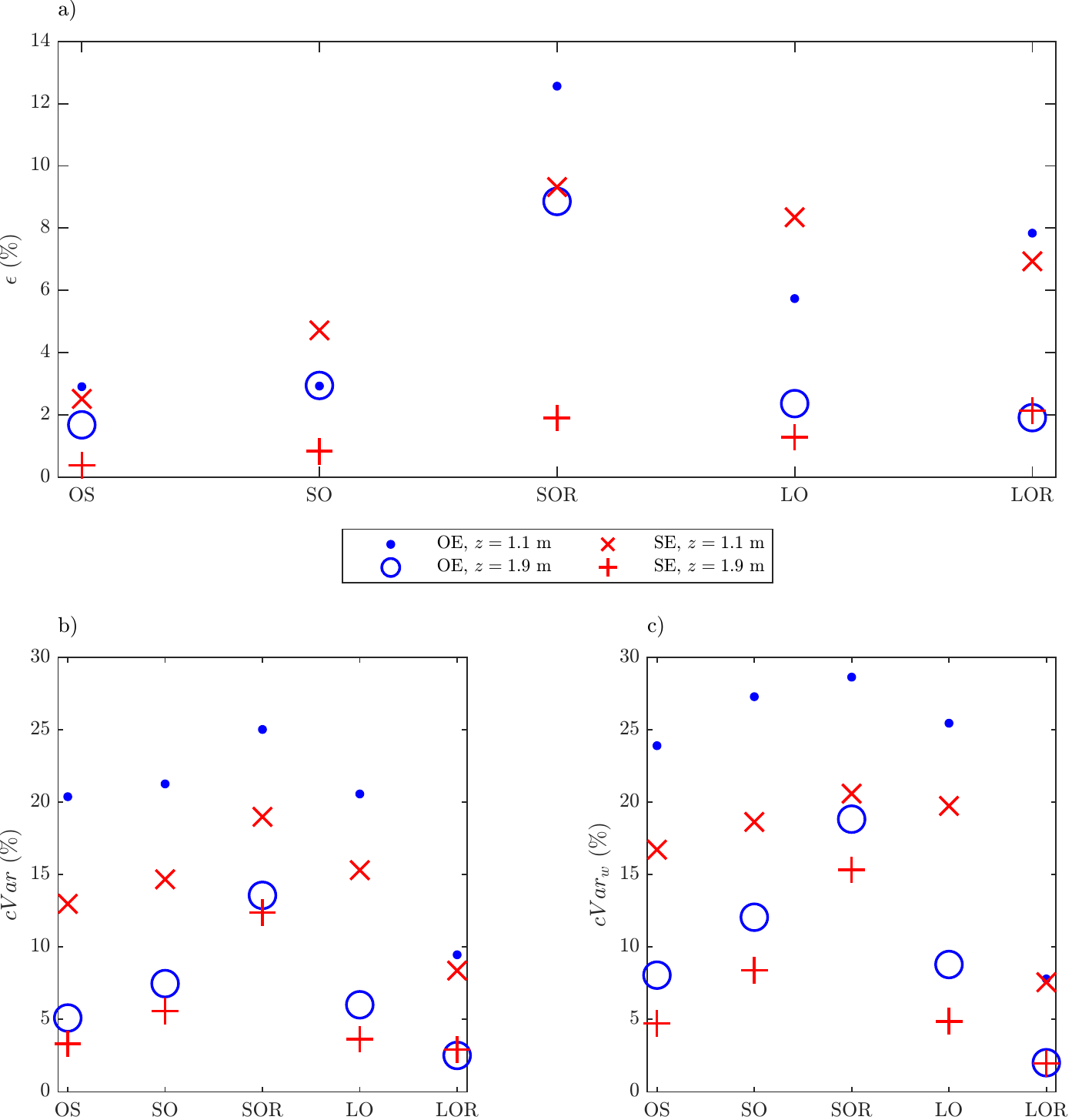}
    \caption{Accuracy of the \co measurements at the walls at $z=1.1$\,m and $z=1.9$\,m for the different vent set-ups. Results for both the opposite-ended and single-ended configurations are shown. $\epsilon$ is the error in the predicted mean when considering the \co concentration near the walls (defined as within 0.2\,m of a wall). $cVar$ is coefficient of variation of the \co concentration over the whole room at a given height and $cVar_{w}$ quantifies the coefficient of variation at the same height considering measurements near the walls only.}
    \label{fig:FRCO2Wall}
\end{figure}

\Cref{fig:FRCO2Wall} presents the scale of the expected errors in the predicted mean using wall measurements, $\epsilon$, and the coefficient of variations found using the measurement over the entire height $cVar$ and at the walls only $cVar_{w}$ for all of the scenarios examined. This is shown for the two heights which show the largest differences in \S\ref{sec:Configsensor}: $z=1.1$\,m and $z=1.9$\,m. As shown previously, the error and coefficients of variations are larger near the breathing zone, which can be expected. Agreement between wall measurements and those over the entire plane are closer higher up within the classroom for all cases, with errors in the predicted mean ranging from less than 1\% to 4\% (coefficients of variation varying between 1\% and 8\% in the domain and 1\% and 12\% at the walls). \rev{This is small compared to the uncertainty associated with \co measurements which is typically taken to be $\pm 50$\,ppm or 3\% of the reading.}  All parameters show distinct peaks at 1.9\,m in the opposite-ended configuration for the smaller opening scenarios (SO, SOR) which are found to have a larger spreads in the \co distributions therefore leading to higher errors and coefficient of variations. In the breathing zone ($z=1.1$\,m), trends are not as clear, the error $\epsilon$ ranges from 1\% to 12\% over the different set-ups, with the single-ended configuration leading to better results (and a smaller error) in all but 2 cases (SO and LO). In both ventilation configurations, the maximum error is reached for the case smaller openings and vent area ratio (SOR). At $z=1.1$\,m, the coefficients of variation vary between 9\% and 25\% in the domain and 9\% and 29\% at the walls, with consistently higher values for the opposite-ended configuration. Here, a significant decrease in the coefficient of variation is visible for the simulation with larger openings and vent area ratio (LOR), which is explained by the increased recirculation and mixing in the room observed in \Cref{fig:Streamlines2}\unskip d).

Across the different simulations considered here, wall measurements are found to be more accurate at points higher in the room. The \co distribution in the breathing zone displays larger variations for all set-ups and would be hard to measure accurately with a limited number of sensors. Despite significant changes in the ventilation pattern created by changes in the vent areas and ratio, conclusions drawn for the original vent set-up hold up for the different vents considered. Larger variations of \co are observed using the opposite-ended configuration and wall measurements taken above the breathing zone and below the high-level vent are found to be the most accurate to represent the distribution at that height. 

\section{Inference of ventilation rates}\label{sec:ventratecalc}

Following the simulations presented under steady state, an estimate of the ventilation rate $Q_E$ can be determined by rearranging \cref{eq:CO2pred} based on the \co measurement, $C$, at any point within the room, giving 
\begin{equation}
        Q_E = \frac{N \, G}{C-C_a} \;. \label{eq:QE}
\end{equation}
The error within the estimated ventilation rate $Q_E$ is therefore determined by the inaccuracies in the measured or estimated values of the other parameters; namely the source term (here expressed as a product of the occupancy, $N$, and the mean production rate per person, $G$), and the difference between the measured carbon dioxide, $C$ and the ambient \co\unskip, $C_a$. We aim to establish how the uncertainty induced by spatial variations of measured \co\unskip, $C$, compare to the magnitude of the uncertainties due to the other three parameters. For instance, although 32 occupants is a standard number for UK classrooms, occupancy levels can be difficult to obtain in practice and so often this number has to be estimated and it might be realistic to assume that $N$ can vary by $\pm$ 5 occupants across classrooms. The individual \co generation rate, $G$ is also extremely variable, depending both on the population and their activity. The value used in this paper ($3.35\times10^{-6}$\,m\ts{3}/s) is obtained from \citet{Persily2017} and is already a population average based on body mass distribution. Although this distribution can have a significant spread, the standard deviation will reduce (statistically by a factor statistically expected to be $\sqrt{N}$) when averaging over $N$ occupants. However, their level of activity remains a vital factor. Herein the activity level is assumed to be 1.4\,met corresponding to the activity levels of a sitting office worker but it can be reasonable to assumed to vary between 1.2\,met (sitting quietly) and 1.6\,met (office worker moving around) in a classroom. This in turn means that $G$ for primary school aged children aged between 6 and 11 could vary between $2.7\times10^{-6}$\,m\ts{3}/s (taking 1.2\,met and young female students to be present) and $4.0\times10^{-6}$\,m\ts{3}/s (taking 1.6\,met and older male students to be present) \cite{Persily2017}. Finally, there is also an uncertainty associated with the outdoor \co level $C_a$ used to estimate the excess \co concentration, which can be assumed to reasonably vary by $\pm 50\,$ppm. 

\begin{figure}[h!]
    \centering
    \includegraphics[width = \textwidth]{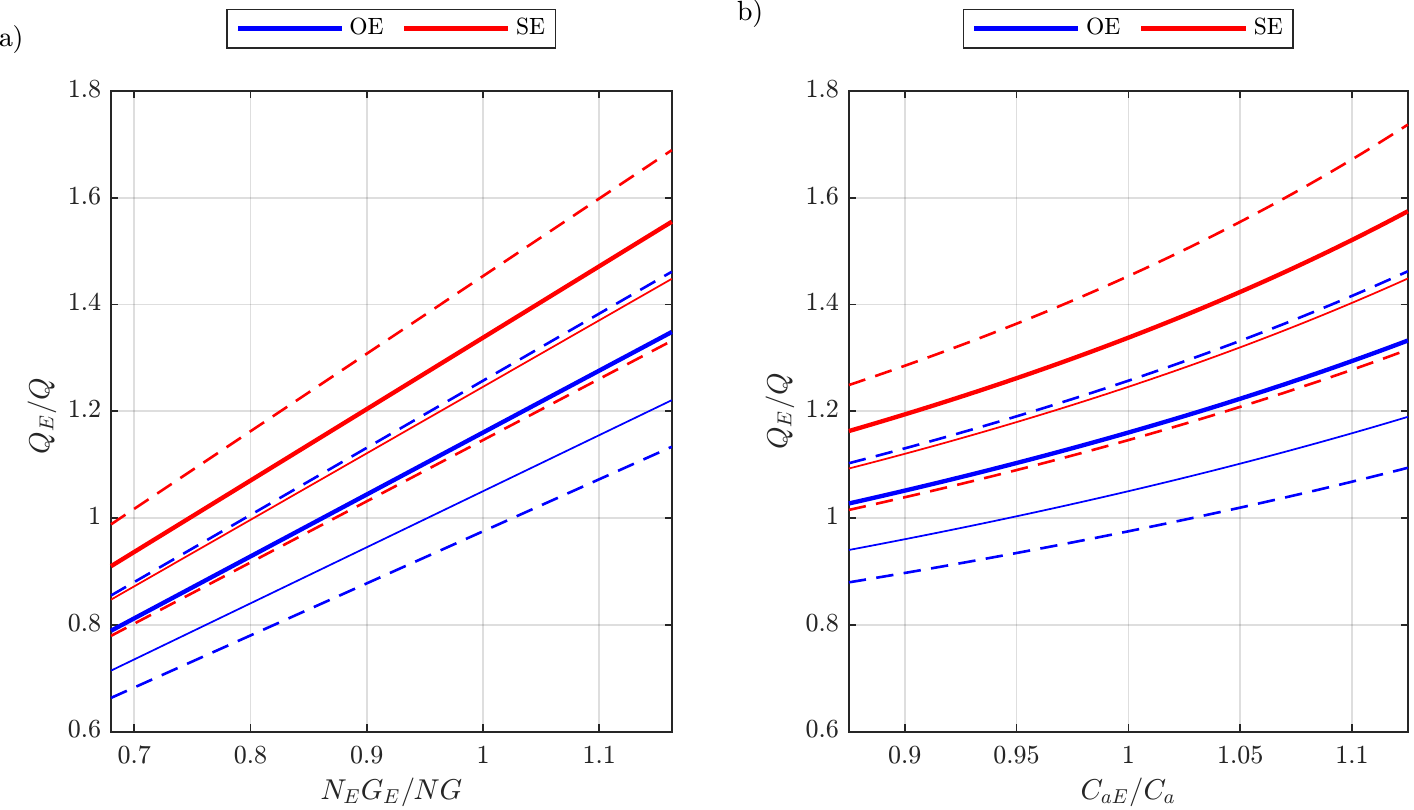}
    \caption{Sensitivity of the estimated ventilation rate to: a) the normalised source term $N_EG_E/NG$ and b) the normalised outdoor \co concentration $C_{aE}/C_a$. $NG$ and $C_a$ are the respective values of the source term and ambient \co concentration used in the numerical simulations and defined in \S\ref{sec:method}. For each case, the estimated ventilation rate $Q_E$ is normalised by the flow rate $Q$ achieved in numerical simulation for the original vent set-up (OS), either in the opposite-ended (OE) or single-ended (SE) configuration (given in \Cref{tab:Config_results}). Thick solid curves show the ventilation rate obtained by using the median \co concentration achieved in each configuration, thin solid curves show the estimates when taking the room averaged \co concentration and the dashed curves show the rate estimated by using the concentrations corresponding either to the 1\ts{st} or 3\ts{rd} quartiles of the distribution (as plotted in \Cref{fig:Configpdf}\unskip a). For each parameter, the variation in the calculated flow rate is shown for the expected range of variations defined in \S\ref{sec:ventratecalc}. }
    \label{fig:sensitivityFR}
\end{figure}

The distribution of the estimated flow rate, $Q_E$ (based on the distribution of measurement points of $C(x,y,z)$ obtained from the simulations), is illustrated in \Cref{fig:sensitivityFR} by curves representing the mean and the quartiles in the distribution of the normalised metric $Q_E/Q$, where $Q$ is actual ventilation flow rate measured within the simulations. Along the horizontal axes within \Cref{fig:sensitivityFR}, the other three parameters are varied to ultimately provide an illustration of the sensitivity of ventilation rates inferred from point measurements, i.e. $Q_E/Q$, relative to other sensitivities associated with classroom operation. In \Cref{fig:sensitivityFR}\unskip a) the estimated source term $N_EG_E$ is varied and normalised by the source term within the simulations, i.e. $NG$; in \Cref{fig:sensitivityFR}\unskip b) the estimated value of the outdoor concentration $Ca_E$ is varied and normalised by $C_a$, the outdoor concentration defined in \S\ref{sec:method}. The \co concentrations measured (at all locations within the room) are taken from the original vent set-up simulations (OS). Within \Cref{fig:sensitivityFR} thin solid curves mark the data of the estimated ventilation rate taking the room averaged \co concentration, i.e. $C=\langle C \rangle$ within \cref{eq:QE}; by construction, these thin curves pass through the points $(1,\eta)$ with the contaminant removal efficiency being $\eta=1.05$ for the opposite-ended configuration, and $\eta=1.24$ for the single-ended. Thick solid curves within \Cref{fig:sensitivityFR} show the estimated ventilation rate using the median \co concentration and dashed curves mark the estimates using the \co concentrations corresponding to the 1\ts{st} and 3\ts{rd} quartiles of the distributions (see \Cref{fig:Configpdf}\unskip a) for histograms of these distributions). All remaining parameters are kept constant and equal to the values described in \S\ref{sec:method}. 

\Cref{fig:sensitivityFR} shows that the estimation of the ventilation rate can vary greatly depending on the particular location of the \co concentration used. Considering the interquartile range within the base-case OS, i.e. taking $N_EG_E/NG$ = 1 or $C_{aE}/C_a=1$, variations in the location of the \co measurement lead to estimations of the ventilation rate that vary by around $\pm15\%$ about the median value. However, \Cref{fig:sensitivityFR} also shows that the uncertainty introduced by the other parameters in \cref{eq:QE} also significantly affects estimates of the ventilation rate. This is particularly significant for the source term $N_EG_E$ (\Cref{fig:sensitivityFR}\unskip a) for which, independent variation over the range considered, leads to variations in the estimated ventilation rate varies by up to 70\%. In the case of the outdoor concentration (\Cref{fig:sensitivityFR}\unskip b), independent variation over the range $\pm 50\,$ppm, results in estimations varying by 35\%. These results highlight that even though the location of \co measurements is indeed important when estimating ventilation rates, the accuracy of these estimates depends strongly on other parameters too; parameters that are, all to often, overlooked and, in the case of the source term, are impractical to measure. 

\section{Conclusion} \label{sec:conc}
Significant spatial variations in the \co concentration were observed in all the ventilation set-ups considered in this study, and consideration of their implications have been presented. It was shown that these spatial variations in the \co concentration result in the precise location of measurements being an important factor to consider when inferring metrics concerning the ventilation. However, uncertainties, e.g. in estimating the scale of the source term, have been shown to be at least as important as the choice of measurement location. Moreover, when reporting ventilation metrics, one must consider the most appropriate rate to report; for example, whether one is trying to report metrics of the bulk ventilation rate (i.e. the volume flux of air exchanged with outdoors) or report metrics of the effectiveness of the ventilation for occupants. These differ by a factor, related to the contaminant removal efficiency, which varies considerably depending on the given ventilation configuration. Practically, there remains questions as to how frequently measurements in classrooms might be deemed to approximately represent steady state \co concentrations, this is especially the case when air change rates are low (either due to ventilation rates being low or room volumes being relatively large). If one takes peak \co concentrations to be representative of the steady value, as has been the case in some previous field studies \citep[e.g.][]{Haverinen-Shaughnessy2011}, then the ventilation provision is likely to be inherently overestimated. 

In the configurations presented, the spatial distribution of \co and resulting contaminant removal efficiency was shown to depend both on the relative horizontal position of the vents as well as the ratio of the vent areas. When the low-level inlet, and high-level outlet, vents were positioned at one end of the classroom (single-ended configuration) the ventilation strategy was shown to be most efficient, consistently leading to lower \co concentrations in the classroom. Changes to the ratio of the vent areas also led to changes in the mixing induced by the fountain created at the low-level vent. This highlights that in addition to the effective area, the relative position of the vents and the ratio of their areas need to be carefully considered to determine the effectiveness of the ventilating flow established, and the resulting \co concentration in spaces designed to promote buoyancy driven ventilation.

Relevant to the choice of location for \co measurements to be made, herein the variations of \co were shown to be highest within the breathing zone ($1\leq z\leq1.5$\,m) with the spread in the observed concentration decreasing at greater heights due to the vertical displacement flow established. Measurements near the walls (within 0.2\,m) were compared to the distribution found over the whole room at each height. This is pertinent in classrooms, as sensors are often placed out of the way and fixed to walls. This comparison shows that, for ventilation strategies which promote low-level inflows and high-level outflows, sensors placed near walls can be useful to predict the mean \co at that height, provided they are above the breathing zone and below the classroom ventilation outlet. \co measurements should however be used with caution when used to infer ventilation rates, even if they are broadly representative of the mean \co\unskip. In the single-ended configuration, for instance, where the contaminant removal efficiency is superior to unity (unity being the well-mixed efficiency), calculating the ventilation flow rate from the room averaged \co would lead to overestimating the actual ventilation supply by 25\%. The robustness of these results was investigated by considering a range of scenarios in which the ventilating flow rate was varied either by changing the opening areas or the ratio of their areas. As expected, a lower flow rate leads to higher \co concentrations, but, crucially, also lead to higher variations in the concentration within the classroom. 

The deployment of CFD simulations, with boundary conditions far removed from the room inlet and outlet vents, enabled the simulation of a naturally ventilated classroom in which the flow rate is set naturally by the heat input and not imposed at the vents. This allowed for the independent calculation of the discharge coefficients at each vent which is often challenging to achieve experimentally (see Appendix \ref{sec:CdCalc}). We hope this might inspire further study since parameterisations of the losses at the vents are key in the design of natural ventilation \citep{CIBSE2005}. \rev{In particular, it would be interesting to determine how the results presented herein are impacted when vents are placed vertically on walls, as is typical of windows.} It must also be noted that \rev{this study assumed convection to be the dominant mechanism of heat transfer and as such the effects of thermal radiation and conduction were not explicitly modelled. Radiation acts to redistribute heat, from warmer to cooler surfaces, within rooms and conduction typically lessens the total net heat load (via heat losses through the building fabric). The inclusion of these mechanisms is likely to impact the vertical temperature distribution, potentially weakening the thermal stratification and lowering any fluid velocities associated with the buoyancy due to temperature differences. Thus}, a useful extension of this work would be to include these effects to determine whether they are expected to significantly affect the findings presented herein, although the computational cost of doing so should not be underestimated.

This work shows that considerable variations in \co can be expected even in idealised naturally ventilated classrooms, even in the limiting case that the heat loads are well distributed. Predictions of concentrations based on the well-mixed assumption should be used with caution, ideally alongside account of the expected variations. If the goal is to estimate the overall ventilation rate, \co measurements taken above the breathing zone were shown to be most accurate for the configurations examined. Irrespective, when metrics of the ventilation are being inferred from point measurements of the \co concentration, the particular measurement location has been highlighted to be just one important factor to consider, and is often not the most important consideration.


\clearpage


\section*{Acknowledgements}
We gratefully acknowledge Fred Mendonça and his team at OpenCFD for providing support and advice on setting up the numerical simulations. CVMV was funded via the Imperial College London Centres for Doctoral Training in Fluid Dynamics Across Scales [grant EP/L016230/1] and the School Air quality Monitoring for Health and Education SAMHE project [grant EP/W001411/1]. Computational resources were provided by the UK Turbulence Consortium [grant EP/R029326/1].

\section*{Competing interests}

The authors declare no competing interests.

\begin{appendices}
\rev{\section{Well-mixed predictions}\label{ap:wellmixed}
The limiting case that the air within the classroom is perfectly mixed and all properties of the air are uniformly distributed throughout the volume, the so-called `well-mixed' approximation, provides predictions for the room averaged temperature $T_c$, \co concentration $C_c$ and ventilation flow rate $Q_c$. These predictions are based on the work of \citet{Gladstone2001} who used a distributed heat source. Such predictions, by their very nature, are unaffected by the horizontal locations of the ventilation openings and are identical for the opposite-ended and single-ended configurations. The expected volume flow rate is 
\begin{equation}
    Q_c = A^{*2/3}\left(\frac{g \, \alpha}{\rho_a \, c_p}\right)^{1/3}W_c^{1/3}H_c^{1/3}\;,
    \label{eq:Qpred}
\end{equation}
where $g$ is the gravitational acceleration, $\alpha$ the thermal expansion coefficient, $\rho_a$ the ambient density, $c_p$ the specific heat capacity, $W_c$ the heat input to the room and $H_c$ is the classroom height. The different parameters and their respective values are summarised in \Cref{tab:SetupInput}. The effective vent area $A^*$ characterises both vent areas and the effect of flow contraction at the openings and is defined as 
\begin{equation}
    A^* = \frac{\sqrt{2} \; c_l \, A_l \, c_h \, A_h}{\sqrt{(c_l \, A_l)^2 + (c_h \, A_h)^2}}\,,
    \label{eq:A*}
\end{equation}
where $A_l$ and $A_h$ are the bottom and top vent areas respectively. The discharge coefficients $c_{l}$ and $c_{h}$ are the discharge coefficients at the low- and high-level vents, their calculation is detailed in Appendix \ref{sec:CdCalc}. The resulting room temperature $T_c$ is given by
\begin{equation}
    T_c = T_a + W_c^{2/3}\left(\rho_a \, c_p \, A^*\right)^{-2/3}\left( g \, \alpha , H_c\right)^{-1/3} \;,
    \label{eq:Tpred}
\end{equation}
where $T_a$ is the ambient temperature. The steady state classroom \co concentration is given by
\begin{equation}
    C_c = \frac{N \, G}{Q_C}+C_a \;,
    \label{eq:CO2pred}
\end{equation}
with $N$ the number of occupants, $G$ their respective \co generation rate and $C_a$ the ambient \co concentration, typically expressed in ppm.}

\section{Independent calculations of the discharge coefficient at each vent}\label{sec:CdCalc}
Theoretical predictions and experimental data evidence the need to characterise the energy losses at the vents. As is standard, we parameterise these losses via discharge coefficients. From the simulated data, the effective area can be evaluated from 
\begin{equation}
    A^*= \frac{Q}{(\int_0^{H_c}g'dz)^{1/2}}\, ,
    \label{eq:AstarCFD}
\end{equation}
where the ventilation flow rate, $Q$, and the buoyancy integral, $\int_0^{H_c}g'dz$ can be determined from the CFD results. Rearranging \cref{eq:A*} and assuming $c_l = c_h = c_d$ as done typically, the discharge coefficient can be determined as
\begin{equation}
    c_d = \frac{A^*}{\sqrt{2}A_lA_h}\sqrt{A_l^2+A_h^2}\, .
\end{equation}
\subsection{Changes in the discharge coefficient with configuration}
The value for both configurations is given in \Cref{tab:VentLosses}. In both cases it is close to 0.7, and so lies within the experimentally determined range of $0.6 \leq c_d \leq 1.0$ \citep{Heiselberg2001,Gladstone2001}.

The assumption of an identical discharge coefficient at each vent can be relaxed by assessing numerically the neutral pressure level $z_{npl}$, the height at which the pressure in the room is equal to the ambient pressure. Then, the ratio 
\begin{equation}
    R^* = \frac{A_lc_{l}}{A_hc_{h}}\, ,
    \label{eq:Rstar}
\end{equation} 
can be found, following the work of \cite{Connick2013} and \cite{Connick2020}, from
\begin{equation}
      R^{*} = \sqrt{\frac{H_C}{z_{npl}}-1}   \,.
\end{equation}
Rearranging \cref{eq:A*}, the discharge coefficient at the low-level vent can be determined from: 
\begin{equation}
    c_l = \frac{A^*}{A_l}\sqrt{\frac{R^{*2}+1}{2}}\, ,
\end{equation}
and $c_h$ is obtained by rearranging \cref{eq:Rstar}. The results for both configurations are summarised in \Cref{tab:VentLosses}, alongside the discharge coefficients $c_d$ obtained when $c_h$ is assumed to be the same as $c_l$. The table also gives the Reynolds number, calculated at each vent with area $A$: 
\begin{equation}
    Re = \frac{Q}{\nu A^{1/2}} \, , \label{eq:Re}
\end{equation}
where the kinematic viscosity is $\nu = 1.37 \times 10^{-5}  $\, m\ts{2}/s, and the Froude number: 
\begin{equation}
    Fr = \frac{Q^{3/2}}{F_s^{1/2}A^{5/4}} \, , \label{eq:Fr}
\end{equation}
where $F_s$ is the source buoyancy flux, found from the distributed heat input with $F_s = \alpha\,g\,W_C/(\rho\, c_p)$. We note that \cref{eq:Re} and \cref{eq:Fr} arise upon taking the characteristic scale of the momentum flux at the vent to be $M = Q^2 / A$. At the high-level vent $Fr_h = Fr$ and the low-level vent $Fr_l = -Fr$ due to the opposing buoyancy force. 

\begin{table}[h!]
    \centering
     \begin{tabular}{c|c|c|c|c|c|c|c|c|}
          \cline{2-9} 
          & $A^*\, (\textrm{m}^2)$ & $c_d$ & $c_l$ & $c_h$ & $Re_l$ & $Re_h$ & $Fr_l$ & $Fr_h$\\ 
          \hline 
          \multicolumn{1}{|c|}{Opposite-ended simulation} & 0.18  & 0.69 & 0.52 &  0.77 &  2.75$\times 10^4$ &  3.89$\times 10^4$ & -0.88 & 2.10\\ 
          \multicolumn{1}{|c|}{Single-ended simulation} & 0.18  & 0.71 & 0.52 &  0.80 &  2.78$\times 10^4$ &  3.93$\times 10^4$ & -0.90 & 2.13\\ 
          \hline 
     \end{tabular}
    \caption{Effective area $A^*$ and discharge coefficients obtained numerically for each ventilation configuration. $c_d$ is obtained by assuming the loss coefficient is the same at each vent, $c_h$ and $c_l$ give the high- and low-level vent discharge coefficient respectively after considering the neutral pressure level. $Re_l$ and $Re_l$ give the Reynolds number, $Fr_l$ and $Fr_h$ the Froude number at the low- and high-level vents respectively.}
    \label{tab:VentLosses}
\end{table}

\Cref{tab:VentLosses} shows that the losses due to contraction at the low-level vent are more significant than at the high-level vent for both ventilation configurations. At the low-level vent, both the Reynolds number and the magnitude of the Froude numbers are lower than at the high-level vent. However, the flow at all vents remains sufficient to be regarded as high-Reynolds number, i.e. the flow is expected to be independent of Reynolds numbers \citep{Etheridge2011}. Given this, and the fact that at both low- and high-level vents the flow experiences both a significant contraction then expansion, our findings may be suggestive that change in sign of the Froude number, i.e. the flow beyond the low-level vent forms a fountain and beyond the high-level vent a plume, plays a significant role in determining the losses at the vent. We note that far less is known about the dependence of the losses at vents due to buoyancy effects, as characterised by the Froude number, than Reynolds number --- this identifies an outstanding challenge. 

In addition, apparent increases in the discharge coefficient at the high-level vent, $c_h$, are evident for the single-ended configuration --- the reasons for this are not clear. The calculation of specific discharge coefficients at each vent requires pressure differences to be measured, this is challenging experimentally -- something numerical simulations are well placed to address. 

\subsection{Changes in the discharge coefficient with vent area}\label{app:CDventarea}
The discharge coefficients, Reynolds numbers, and Froude numbers are also calculated for the simulations discussed in \S \ref{sec:FlowRate}, and are presented in \Cref{tab:SimListFlowRate}. This shows that the discharge coefficient at the low-level vent, $c_l$, remains low, relative to $c_h$, in all cases; supporting the suggestion that buoyancy effects across the vents may be important in determining the losses. Moreover, $c_l$, remains equally low for both the single- and opposite-ended configuration in all of the cases examined. This does not remain true for $c_h$, where a lower value is often found in the opposite-ended configuration. Changes due to variations in the vent areas are most visible at the low-level. Whilst $c_h$ remains within the range 0.74--0.8 for all the cases considered, $c_l$ varies more widely (0.46--0.62), reaching a minimum for the case with smaller openings and vent area ratio (SOR) and a maximum for the set-up with larger openings and vent area ratio (LOR).
\begin{table}[h!]
    \centering
     \begin{tabular}{|c|c|c|c|c|c|c|c|c|c|}
     \hline 
     Case  & Vents & $A_h/A_l$ & $A^*\, (\textrm{m}^2)$ &$c_l$ & $c_h$ & $Re_l$ & $Re_h$ & $Fr_l$ & $Fr_h$ \\ 
     \hline 
     \multirow{2}{*}{OS} & OE & \multirow{2}{*}{0.50} & 0.18 &0.52 &0.77& 2.8$\times 10^4$ &  3.9$\times 10^4$ & -0.88 & 2.10 \\ 
      & SE &   & 0.18 &0.52 &0.80 & 2.8$\times 10^4$ &  3.9$\times 10^4$ & -0.90 & 2.13   \\ 
     \hline 
     \multirow{2}{*}{SO} & OE & \multirow{2}{*}{0.50} & 0.09 &0.55 &0.78& 2.5$\times 10^4$ &  3.5$\times 10^4$ & -1.09 & 2.58 \\ 
      & SE &   & 0.09 &0.55 &0.78 & 2.5$\times 10^4$ &  3.5$\times 10^4$ & -1.09 & 2.58   \\ 
     \hline 
     \multirow{2}{*}{SOR} & OE & \multirow{2}{*}{0.25} & 0.10 &0.46 &0.77& 1.9$\times 10^4$ &  3.8$\times 10^4$ & -0.51 & 2.86 \\ 
      & SE &   & 0.10 &0.46 &0.77 & 1.9$\times 10^4$ &  3.8$\times 10^4$ & -0.51 & 2.87   \\ 
     \hline 
     \multirow{2}{*}{LO} & OE & \multirow{2}{*}{0.50} & 0.34 &0.51 &0.74& 3.0$\times 10^4$ &  4.2$\times 10^4$ & -0.70 & 1.68 \\ 
      & SE &   & 0.35 &0.51 &0.76 & 3.0$\times 10^4$ &  4.2$\times 10^4$ & -0.71 & 1.69   \\ 
     \hline 
     \multirow{2}{*}{LOR} & OE & \multirow{2}{*}{2.00} & 0.32 &0.62 &0.74& 4.1$\times 10^4$ &  2.9$\times 10^4$ & -1.58 & 0.66 \\ 
      & SE &   & 0.33 &0.62 &0.80 & 4.1$\times 10^4$ &  2.9$\times 10^4$ & -1.61 & 0.68   \\ 
     \hline 
     \end{tabular} 
    \caption{Effective area $A^*$ and discharge coefficients at each vent calculated for each set-up. $Re_l$ and $Re_h$ are the Reynolds numbers, and $Fr_l$ and $Fr_h$ are the Froude numbers at the low- and high-level vents respectively.}
    \label{tab:SimListFlowRate}
\end{table}
\end{appendices}

\bibliographystyle{elsarticle-num-names}
\bibliography{bibl_USE.bib}
\end{document}